\documentclass[article]{aa} 
\usepackage{graphicx}
  
\usepackage{tabls}
\usepackage{array}
\usepackage{longtable}

\begin{document}

\title{ XMM-Newton observations of Seyfert galaxies from the Palomar spectroscopic survey:  the X-ray absorption distribution}
\titlerunning{X-ray absorption in the local universe}
\authorrunning{A. Akylas and I. Georgantopoulos}
\author{A. Akylas and I. Georgantopoulos}
\offprints{A. Akylas}

  \institute{Institute of Astronomy \& Astrophysics, National Observatory 
 of Athens, I.Metaxa \& B. Pavlou, Penteli, 15236, Athens, Greece \\
     }   
  \date{}

\def\etal{{et al.\ }}
\def\etal{{et al.\ }}
\def\ee{$e^\pm$\ }
\def\g{$\Gamma$}
\def\x2{$\chi^{2}$}
\def\zm{z_{\rm max}}
\def\nh{N_{\rm H}}
\def\nlo{\langle n L\rangle_0}
\def\af{A_{\rm Fe}}  
\def\taut{\tau_{\rm T}}
\def\ec{E_{\rm c}}
\def\ginga{{\it Ginga}}
\def\asca{{\it ASCA}}
\def\exosat{{\it EXOSAT}}
\def\einstein{{\it EINSTEIN}}
\def\xmm{{\it XMM-Newton}~}
\def\chandra{{\it Chandra}~}
\def\integral{{\it INTEGRAL}~}
\def\luv{L_{\rm UV}}
\def\lxg{L_{{\rm X}\gamma}}
\def\asc{a_{\rm sc}}
\def\granat{{\it GRANAT}}
\def\lunits{$\rm erg~s^{-1}$}
\def\funits{$\rm erg~cm^{-2}~s^{-1}$}
\def\cunits{$\rm cm^{-2}$}

\begin{abstract}
{We present {\it XMM-Newton} spectral analysis of all 38 Seyfert galaxies from the Palomar spectroscopic 
sample of galaxies. These are found at distances of up to 67 Mpc and cover the absorbed 2-10 keV luminosity range   
$\sim 10^{38}-10^{43}$ \lunits. 
Our aim is to determine the distribution of the X-ray absorption in the local Universe. 
Three of these are Compton-thick  with column densities just above $ 10^{24}$ {\cunits} and 
high equivalent width FeK$_{\alpha}$ lines ($>700$ eV). Five more sources have low values of the X-ray to 
[OIII] flux ratio 
suggesting that they could be associated with obscured nuclei. 
Their individual spectra show neither high absorbing columns nor flat spectral indices. 
However, their stacked spectrum reveals an 
absorbing column density of $\rm N_H\sim 10^{23}$ \cunits. Therefore the fraction of absorbed
sources ($>10^{22}$ \cunits) could be as high as $55\pm12$ \%.   
A number of Seyfert-2 appear to host  unabsorbed nuclei.  These are associated with 
low-luminosity sources $\rm L_X < 3\times 10^{41}$ \lunits.
Their stacked spectrum again shows no absorption while inspection of the \chandra images, 
where available, shows that contamination from nearby sources does not affect the {\it XMM-Newton} spectra
in most cases.   Nevertheless, such low luminosity sources are not contributing significantly to the X-ray background flux.   
When we consider only the brighter, $>10^{41}$ \lunits, 21 sources, 
we find that the fraction of absorbed nuclei rises to $75\pm19 $ \% while that of Compton-thick 
sources to 15-20\%.   
The fraction of Compton-thick AGN is lower than  that predicted by the X-ray background synthesis model
in the same luminosity and redshift range.}
\end{abstract}

\maketitle

	   	
\section{Introduction}

The moderate to high redshift Universe has been 
probed at unparallelled depth with the most sensitive  
observations performed  at X-ray wavelengths in the 
{\it Chandra} Deep fields. The {\it Chandra} 2Ms 
observations  (Alexander et al. 2003,  Luo et al. 2008) 
resolved about 80 per cent of the extragalactic X-ray light 
in the hard 2-10 keV band (see Brandt \& Hasinger 2005 for a  review). 
These deep surveys find a sky density of 5000 sources per square 
degree, the vast majority of  which are found to be AGN through 
optical spectroscopy (e.g. Barger et al. 2003). In contrast, the 
optical surveys for QSOs (e.g. the COMBO-17 survey) reach only a 
surface density about an order  of magnitude lower (e.g. Wolf et al. 2003).
This clearly demonstrates the power of X-ray surveys  for detecting AGN.   
This is because hard X-rays can penetrate large amounts of gas without 
suffering from significant absorption.  Indeed detailed spectral analysis 
on X-ray selected AGN reveals large amount of obscuration 
(e.g. Akylas et al. 2006, Tozzi et al. 2006, Georgantopoulos et al. 2007). 
 In particular, about two 
thirds of the X-ray sources, over all luminosities, present column 
densities higher than $10^{22}$ \cunits. These high absorbing columns 
are believed to originate in a molecular torus surrounding the nucleus.   

However, even the efficient 2-10 keV X-ray surveys may be missing a 
fraction of highly obscured sources. This is because at very high 
obscuring column densities ($>10^{24}$ \cunits, corresponding to an 
optical reddening of $\rm A_V>100$), the X-ray photons with 
energies between 2 and 10 keV are absorbed. These are the Compton-thick 
AGN (see Comastri 2004 for a review) where the  
Compton scattering on the bound electrons becomes significant. 
Despite the fact that Compton-thick AGN are abundant in our 
vicinity (e.g. NGC1068, Circinus), only a few tens of Compton-thick sources 
have been identified from X-ray data (Comastri 2004). Although the 
population of Compton-thick sources remains elusive there is concrete 
evidence for its presence. The X-ray background synthesis models can 
explain the peak of the X-ray background at 30-40 keV, where most of 
its energy density lies, (Frontera et al. 2007, Churazov et al. 2007) 
only by invoking a large number of Compton-thick AGN 
(Gilli, Comastri \& Hasinger 2007).  Additional evidence for the presence of 
a Compton-thick population comes from the directly measured space density of 
black holes in the local Universe. It is found that this space density is a 
factor of two higher than that predicted from the X-ray luminosity function 
(Marconi et al. 2004). This immediately suggests that the X-ray luminosity 
function is missing an appreciable number of  obscured AGN.

In recent years there have been many efforts to uncover heavily obscured and 
in particular Compton-thick AGN in the local Universe 
by examining IR or optically selected, [OIII], AGN samples. 
This is because both the IR and the narrow-line region originate beyond the 
obscuring region and thus represent an isotropic property of the AGN. 
Risaliti et al. (1999) examine the X-ray properties of a large sample of 
[OIII] selected Seyfert-2 galaxies whose X-ray spectra were available in 
the literature. They find a large fraction of Compton-thick sources  
(over half of their sample). Their estimates are complemented
by more recent {\it \xmm} observations  of local AGN samples 
(Cappi et al. 2006, Panessa et al. 2006, Guainazzi et al. 2005). 
All these authors also claim a large  Compton-thick AGN fraction exceeding 
30 per cent of the Seyfert-2 population. 
The advent of the {\it SWIFT} and {\it INTEGRAL} missions which carry X-ray 
detectors with imaging capabilities (e.g. Barthelmy et al. 2005, Ubertini et al. 2003) 
in ultra-hard X-rays (15-200 keV)  try to  shed new light on the 
absorption properties of AGN in the local Universe.  
In principle, at these ultra-hard X-rays obscuration should play a 
negligible role, at least up to column densities as high as $10^{25}$ \cunits. 
However, because of the limited effective area the above surveys can 
provide X-ray samples, down to very bright fluxes $10^{-11}$ \funits, with 
limited quality spectra. Again  \xmm observations are often required to 
determine the column density in each source. Interestingly, these surveys 
find only a limited number of Compton-thick sources (Markwardt et al. 2005, 
Bassani et al. 2006, Malizia et al. 2007, Ajello 2008, 
Winter et al. 2008,  Tueller et al. 2008,  Sazonov et al. 2008).  

Here, we present \xmm observations of {\it all} 38  Seyfert 
galaxies in the Palomar spectroscopic sample of nearby galaxies 
(Ho et al. 1997). This is the largest complete optically selected 
AGN sample in the local Universe analyzed so far. 
23 of the Seyfert galaxies presented here have already been 
discussed in previous works (e.g. Cappi et al. 2006). For 5 of 
them newer \xmm observations are available and are presented here.
The current work should provide the most unbiased census of the AGN column 
density distribution at low redshifts and luminosities.  

\section{The sample}{\label{thesample}} 

The Seyfert sample used in this study is derived from the 
Palomar optical spectroscopic survey of nearby galaxies 
(Ho, Fillipenko, \& Sargent 1995). This survey has taken 
high quality spectra of 486 bright ($B_T<12.5$ mag), 
northern ($\delta>0^{\circ}$) galaxies selected from the Revised Shapley-Ames 
Catalogue of Bright Galaxies (RSAC, Sandage \& Tammann 1979) and produced 
a comprehensive and homogeneous catalogue of nearby Seyfert  galaxies. 
The catalogue is 100\% complete to $B_T<12.0$ mag and 80\% complete to 
$B_T<12.5$ mag (Sandage, Temmann \& Yahil 1981). 

For the purpose of this work we consider all the Seyfert galaxies 
from the Palomar survey. Sources  lying in-between the Seyfert-Liner or 
the Seyfert-Transient boundary have been excluded.  
Furthermore seven Seyfert galaxies (i.e. NGC1068, NGC1358, NGC1667, NGC2639, 
NGC3185, NGC4235, NGC5548), which have been included in the Palomar 
survey for various reasons (see Ho et al. 1995), even though  
they did not satisfy the survey selection criteria, are also  
excluded.

There are 40 Seyfert galaxies comprising the optical sample. 
9 sources are classified as type-1  (contains  types 1, 1.2, 1.5) 
and 31 as type-2  (contains types 1.8,1.9,2) Seyfert galaxies. 
However NGC4051, NGC4395 and NGC4639 which have been  initially classified 
as Seyfert 1.2, 1.8  and 1 by Ho et al. (1997) has been re-classified 
as type-1.5, 1 and 1.5 respectively (see Cappi et al. 2006, 
Panessa et al. 2006, Baskin \& Laor 2008).
Moreover NGC185 which is classified as a Seyfert-2 
may not contain an active nucleus since it presents 
line intensity ratios possibly produced by stellar processes 
(Ho \& Ulvestad 2001). 

The main characteristics of these sources, taken from Ho et al. (1997), 
are listed in Table \ref{optical}. 
Some galaxies listed here present $B_T$ fainter than the formal 
limit of the Palomar survey. According to Ho et al. (1995) this 
discrepancy can be attributed to errors in the apparent magnitudes 
given in the RSAC.

\begin{table*}
\centering
\caption{The sample}
\label{optical}
\begin{tabular}{cccccc}
Name & $\alpha$(J2000) & $\delta$(J2000) & $B_T$ (mag) & D(Mpc) & Class  \\
					
(1)& (2)  &(3) & (4) & (5)& (6)  \\
\hline
NGC 0185  & 00 38 57.40  & +48 20 14.4   &10.10 &0.7 &S2 \\
NGC 0676  & 01 48 57.38  & +05 54 25.70  &10.50 &19.5&S2: \\
NGC 1058 & 02 43 30.24 & +37 20 27.20   &11.83 &9.1 &S2 \\
NGC 1167 & 03 01 42.40 & +35 12 21.00   &13.38 &65.3& S2 \\
NGC 1275 & 03 19 48.16 & +41 30 42.38   &12.64 &70.1& S1.5  \\
NGC 2273 & 06 50 08.71 & +60 50 45.01   &12.55 &28.4& S2\\
NGC 2655 & 08 55  38.84& +78 13 25.20   &10.96 &24.4& S2\\
NGC 3031 & 09 55 33.17 &+69 03 55.06    & 7.89 &1.4& S1.5\\
NGC 3079 & 10 01 58.53 &+55 40 50.10    &11.54 &20.4 & S2\\
NGC 3147 & 10 16 53.27 &+73 24 02.40    &11.43 &40.9& S2 \\
NGC 3227 & 10 23 30.58 & +19 51 53.99   &11.10 &20.6 & S1.5 \\
NGC 3254 & 10 29 19.96 & +29 29 29.60   &12.41 &23.6& S2 \\
NGC 3486 & 11 00 24.10  &+28 58 31.60   &11.05 &7.4 & S2 \\
NGC 3516 & 11 06 47.49 & +72 34 06.80   &12.50 &38.9& S1.2 \\
NGC 3735 & 11 35 57.49  &+70 32 07.70   &12.50 &41.0& S2: \\
NGC 3941 & 11 52 55.42 &+36 59 10.50    &11.25 &18.9& S2: \\
NGC 3976 & 11 55 57.35 &+06 44 57.00    &12.30 &37.7& S2: \\
NGC 3982 & 11 56 28.10 &+55 07 30.50    &11.78 &17.0& S1.9 \\
NGC 4051 & 12 03 09.61 &+44 31 52.80    &11.88 &17.0& S1.2 \\
NGC 4138 & 12 09 29.87 &+43 41 06.00    &12.16 &17.0& S1.9 \\
NGC 4151 & 12 10 32.57 &+39 24 20.63    &11.50 &20.3& S1.5 \\
NGC 4168 & 12 12 17.30  &+13 12 17.9    &12.11 &16.8& S1.9: \\
NGC 4169 & 12 12 18.93 &+29 10 44.00    &13.15 &50.4 & S2 \\
NGC 4258 & 12 18 57.54 &+47 18 14.30    &9.10  &6.8 & S1.9 \\
NGC 4378 & 12 25 18.14 &+04 55 31.60    &12.63 &35.1& S2 \\
NGC 4388 & 12 25 46.70 &+12 39 40.92    &11.76 &16.8& S1.9 \\
NGC 4395 & 12 25 48.93 &+33 32 47.80    &10.64 &3.6 & S1.8 \\
NGC 4472 & 12 29 46.76 &+07 59 59.90    & 9.37 &16.8& S2:: \\
NGC 4477 & 12 30 02.22 &+13 38 11.30    &11.38 &16.8& S2 \\
NGC 4501 & 12 31 59.34 &+14 25 13.40    &10.36 &16.8& S2 \\
NGC 4565 & 12 36 21.07 &+25 59 13.50    &10.42 &9.7& S1.9 \\
NGC 4639 & 12 42 52.51 &+13 15 24.10    &12.24 &16.8& S1 \\
NGC 4698 & 12 48 22.98 &+08 29 14.80    &11.46 &16.8& S2 \\
NGC 4725 & 12 50 26.69 &+25 30 02.30    &10.11 &12.4& S2: \\
NGC 5033 & 13 13 27.52 &+36 35 37.78    &10.75 &18.7& S1.5 \\
NGC 5194 & 13 29 52.37 &+47 11 40.80    &8.96 &7.7& S2 \\
NGC 5273 & 13 42 08.33 &+35 39 15.17    &12.44 &21.3& S1.5 \\
NGC 6951 & 20 37 14.41 &+66 06 19.70    &11.64 &24.1& S2 \\
NGC 7479 & 23 04 56.69 &+12 19 23.20    &11.60 &32.4& S1.9 \\
NGC 7743 & 23 44 21.44 &+09 56 03.60    &12.38 &24.4 & S2 \\
\end{tabular}
\begin{list}{}{}

\item{Column 1: Galaxy name}
\item{Columns 2 \& 3: Optical coordinates}
\item{Column 4: Total apparent $B$ magnitude  taken from Ho et al. 1997}
\item{Column 5: Source distance in Mpc from Ho et al. 1997}
\item{Column 6: Optical classification from Ho et al. 1997. Quality ratings 
are given by ":" and "::" for uncertain and highly uncertain classification.}
\end{list}{}{}

\end{table*}

\section{X-ray Observations}

The X-ray data have been obtained with the EPIC 
(European Photon Imaging Cameras; Str\"{u}der et al. 2001, 
Turner et al. 2001) on board {\xmm}. Thirty sources 
have been recovered from the \xmm archive while the
remaining ten objects (marked with a ''$\star$'' in 
Table \ref{xray}) have been observed by us during the 
Guest Observer program. 

The log of all the \xmm observations is shown 
in Table \ref{xray}. The data have been analysed using the 
Scientific Analysis Software ({\sl SAS v.7.1}). We produce 
event files for both the PN and the MOS observations 
using the {\sl EPCHAIN} 
and {\sl EMCHAIN} tasks of {\sl SAS} respectively. The event 
files are screened for high particle background periods. 
In our analysis we deal only with events corresponding to 
patterns 0-4 for the PN and 0-12 for the MOS instruments.

The source spectra are extracted from circular regions 
with radius of 20 arcsec. This area encircles at least 
the 70 per cent of the all the X-ray photons  
at off-axis angles less than 10 arcmin. A ten times 
larger source-free area is used for the background 
estimation. The response and ancillary files are also 
produced using {\sl SAS} tasks {\sl RMFGEN} and {\sl ARFGEN} 
respectively.

We note that 18 of the \xmm observations presented here, 
coincide with these presented in Cappi et al. (2006). 
However we choose to re-analyze these common data-sets in order to present  
a uniform treatment of the sample. 
 
\begin{table*}
\centering
\caption{Log of the {\xmm} observations}
\label{xray}
\begin{tabular}{ccccccccc}
\hline
\hline
Name & Obs. Date & Obs. ID & \multicolumn{3}{c}{Exposure}  & \multicolumn{3}{c}{Filter} \\ 		
     &          &          &     PN  & MOS1 &  MOS2         & PN      &  MOS1  & MOS2  \\
\hline
NGC 185 &2004-01-09 &0204790301 &   -   & 11393 & 11334 & closed & Medium & Medium \\
NGC 676 &2002-07-14 &0112551501 & 17754 & 21127 & 21127 & Thick  & Thin   & Thin   \\
NGC 1058&2002-02-01 &0112550201 & 12902 & 17019 & 17019 & Medium & Thin   & Thin   \\
NGC 1167$^{\star}$& 2005-08-04& 0301650101 &  9937 & 11448 & 11448 & Thin   & Thin   & Thin   \\ 
NGC 1275&2006-01-29 &0305780101 &119697 &124801 &124832 & Medium & Medium & Medium \\
NGC 2273&2003-09-05 &0140951001 & 11076 & 12709 & 12714 & Medium & Medium & Medium \\
NGC 2655$^{\star}$&2005-09-04 &0301650301 &  9850 & 11564 & 11570 &  Thin  & Thin   & Thin   \\
NGC 3031&2001-04-22 &0111800101 &129550 & 82790 & 83150 & Medium & Medium & Medium \\
NGC 3079&2001-04-13 &0110930201 & 20023 & 24661 & 24663 &  Thin  & Medium & Medium \\
NGC 3147&2006-10-06 &0405020601 & 14963 & 16923 & 16912 & Thin   & Thin   & Thin   \\
NGC 3227&2000-11-28 &0101040301 & 34734 & 37198 & 37201 & Medium & Medium & Medium \\
NGC 3254$^{\star}$&2005-10-31 &0301650401 &  9869 & 11489 & 11481 & Thin   & Thin   & Thin   \\
NGC 3486&2001-05-09 &0112550101 &  9057 &  6398 &  6385 & Medium & Thin   & Thin   \\
NGC 3516&2001-11-09&0107460701 & 12829 & 12901 & 12900 & Thin   & Thin   & Thin   \\
NGC 3735$^{\star}$&2005-09-27 &0301650501 &  9312 & 16466 & 16471 & Thin   & Thin   & Thin   \\
NGC 3941&2001-05-09 &0112551401 &  9389 & 14635 & 14331 & Medium & Thin   & Thin   \\
NGC 3976$^{\star}$&2006-06-16 &0301651801 & 11313 & 13483 & 13598 & Thin   & Thin   & Thin   \\
NGC 3982& 2004-06-15 &0204651201 & 10197 & 11674 & 11679 & Thin   & Thin   & Thin   \\
NGC 4051& 2002-11-22 &0157560101 & 49808 & 51510 & 51520 & Medium & Medium & Medium \\
NGC 4138& 2001-11-26 &0112551201 &  9999 & 14365 & 14365 & Medium & Thin   & Thin   \\
NGC 4151& 2003-05-26 &0143500201 & 18454 & 18602 & 18607 & Medium & Medium & Medium \\
NGC 4168& 2001-12-04 &0112550501 & 18498 & 22864 & 22849 & Medium & Thin   & Thin   \\
NGC 4169$^{\star}$&2006-06-20 &0301651701 & 11068 & 12695 & 12701 & Thin   & Thin   & Thin   \\
NGC 4258&2006-11-17 &0400560301 & 62607 & 64179 & 64184 & Medium & Medium & Medium \\
NGC 4378$^{\star}$&2006-01-08 &0301650801 & 10963 & 12602 & 12604 & Thin   & Thin   & Thin   \\
NGC 4388&2002-12-12 &0110930701 &  8292 & 11666 & 11666 & Thin   & Medium & Medium \\
NGC 4395&2003-11-30 &0142830101 & 10596 & 10942 & 10940 & Medium & Medium & Medium \\
NGC 4472&2004-01-01 &0200130101 & 89503 & 94179 & 94185 & Thin   & Thin   & Thin   \\
NGC 4477&2002-06-08 &0112552101 &  9500 & 13501 & 13527 & Medium & Thin   & Thin   \\
NGC 4501&2002-06-08 &0112550801 &  2885 & 13387 & 13385 & Medium & Thin   & Thin   \\
NGC 4565&2001-07-01 &0112550301 & 10010 & 14261 & 14263 & Medium & Thin   & Thin   \\
NGC 4639&2001-12-16 &0112551001 & 10000 & 14365 & 14265 & Medium & Thin   & Thin   \\
NGC 4698&2001-12-17 &0112551101 & 11755 & 16112 & 16112 & Medium & Thin   & Thin   \\
NGC 4725&2002-06-14 &0112550401 & 13369 & 17244 & 17244 & Medium & Thin   & Thin   \\
NGC 5033&2002-12-18 &0094360501 &  9999 & 11616 & 11614 & Medium & Medium & Medium \\
NGC 5194&2003-01-15 &0303420101 & 19047 & 49944 & 49351 & Thin   & Thin   & Thin   \\
NGC 5273&2002-06-14 &0112551701 & 10392 & 16065 & 16094 & Medium & Thin   & Thin   \\
NGC 6951$^{\star}$&2005-06-05 &0301651401 &  7951 &  9664 &  9669 & Thin   & Thin   & Thin   \\
NGC 7479$^{\star}$&2005-06-05 &0301651201 & 12315 & 15740 & 15750 & Thin   & Thin   & Thin   \\
NGC 7743$^{\star}$&2005-06-15 &0301651001 & 11847 & 13283 & 13348 & Thin   & Thin   & Thin   \\
\end{tabular}                                                
\begin{list}{}{}                                                                                         
\item{Column 1: Name of the Galaxy} 
\item{Column 2: Start Observation date (UTC)}        
\item{Column 3: Observation identifier}             
\item{Columns 3, 4 \& 5: Net exposure time for the EPIC instruments}
\item{Columns 6, 7 \& 8: Applied filter}  
\item{$^{\star}$ Denotes sources observed during our Guest Observer program}
\end{list}
\end{table*} 

\section{X-ray  Spectral Analysis} 

We investigate the X-ray properties of the sources 
in our sample by performing spectral fittings with 
{\sl XSPEC v.12.4} software package.
2 sources are excluded from the X-ray spectral analysis:
the Seyfert-2 galaxy NGC185 for being undetected 
in the X-rays (see also section \ref{thesample}), 
and the Seyfert-1.5 galaxy NGC1275 which belongs to the 
Perseus cluster and whose X-ray image shows that its 
flux is heavily contaminated by diffuse emission.

The X-ray spectra are binned to give a minimum of 15 counts  so 
Gaussian statistics can be applied. 
We fit the PN and the MOS data simultaneously in the 
0.3-10 keV range. However in some cases where a very complex 
behaviour is present we perform the spectral fits only in the 2-10 keV 
band. These latter cases are denoted with an asterisk 
($\star$) in Table \ref{fit}.

The normalization parameters for each instrument are left 
free to vary within 5 per cent in respect to each other 
to account for the remaining calibration uncertainties.

We assume a standard power-law model with two absorption components 
({\sl wa*wa*po} in {\sl XSPEC} notation) to account for the source 
continuum emission. The first absorption column models the Galactic 
absorption. Its fixed values  are obtained from Dickey \& Lockman (1990) 
and are listed in Table \ref{fit}. The second absorption component 
represents the AGN intrinsic absorption and is left as a free parameter 
during the model fitting procedure.
A Gaussian component has also been included to describe the FeK$_{\alpha}$ 
emission line.

When the fitting procedure gives a rejection probability less than 
90 per cent we accept the above "standard model". 
However when this simple parametrization is not sufficient 
to model the whole spectrum additional components are included. 
For example soft-excess emission and reflection are common 
features in the X-ray spectra of Seyfert galaxies and can be 
modeled using additional {\sl XSPEC} models.

In particular we fit a second power-law model, 
with $\Gamma$ fixed to the direct component value, 
to account for the scattered X-ray radiation 
and/or a {\sl Raymond-Smith} to model the 
contribution from diffuse emission in the host galaxy.
A flattening of the spectrum is usually indicative 
of reflected radiation from the backside of the torus.  
The reflected radiation is modelled using 
the {\sl PEXRAV} model (Magdziarz \& Zdziardski, 1995).  
In order to accept the new component we apply the F-test 
criterion. If the addition of the new component significantly 
improves the fit at the 90 per cent confidence level, 
then it is accepted.  
Other characteristics such as ionized features could also be 
considered however once a reasonable fit is obtained 
(i.e. with rejection probability less than 90 per cent) 
we do not include additional components.

The best fit parameters for all the sources are reported 
in Table \ref{fit}. The errors quoted correspond to the 90 per cent 
confidence level for one interesting parameter. We note here that 
some of the sources listed show a rather steep photon index. 
In many cases this happens because of the fixed value of the 
continuum power-law photon index to the photon index of the 
soft component (e.g. NGC1358, NGC3079, NGC3735). 
When these parameters are untied the continuum power-law photon 
index becomes  harder.

18 of the X-ray observations presented here have already been 
shown in Cappi et al. 2006. In most of these the 
results are in agreement.
However some deviations also appear and are discussed below.
In the cases of NGC3486, NGC3079, NGC4051 and NGC4388   
the comparison is not straightforward since we use of a 
different spectral fitting model. When the same model is applied 
as a test,  there is no significant difference in the results. 
In the cases of NGC1058 and NGC4725 our results show a  steeper 
power-low photon index than that presented in Cappi et al. 2006.
However we point out that the results are  consistent within the 90 
per cent confidence level.

\begin{table*}
\centering
\caption{Spectral fits}
\label{fit}
\begin{tabular}{ccccccccccc}
\hline
\hline
Name & $N_{H_{GAL}}$ & $N_H$ & {$\Gamma_{cont}$} & $kT$ & EW$_{FeK}$ & $F_X$ & $L_X$ & $f_{scat}$ & $f_{refl}$ & $\chi^2_{\nu}$ \\
(1)  & (2) & (3) & (4) & (5) & (6) & (7) & (8) & (9) & (10) & (11)  \\
\hline
NGC676  &4.7  & $<0.13$ & $2.10^{+0.50}_{-0.27}$& - & - & 1.12  & 0.05  & - & - & 22.5/17  \\
NGC1058  & 5.4 & $<0.53$ & $3.40^{+2.65}_{-1.69}$&  -  & -  &  0.24  & 0.02  & -  & -  & 31.1/34 \\  
NGC1167  & 9.8 & $0.32^{+0.08}_{-0.41}$ & $2.10^{+0.50}_{-0.27}$ & $0.33^{+0.08}_{-0.04}$  &-  &  0.9  &  0.4  & -  & -  & 22.52/17 \\ 
NGC2273  & 6.4 & $104.70^{+15.90}_{-7.36}$ & $1.84^{+0.12}_{-0.05}$ & $0.77^{+0.08}_{-0.13}$ & $1500^{+303}_{-240}$ & 89.4 &  8.5& 0.01 & 1.35 & 125.1/114  \\
NGC2655  &2.2  & $42.69^{+5.68}_{-5.30}$ & $2.61^{+0.30}_{-0.26}$ & $0.73^{+0.30}_{-0.26}$  &  110.4  & 0.01  & 7.8 &- & - & 240.1/194 \\ 
NGC3031$^{\star}$  &5.6  &$0.05^{+0.02}_{-0.01}$ & $1.91^{+0.02}_{-0.08}$ & $1.10^{+0.02}_{-0.02}$ & $53^{+11}_{-14}$  & 1164.7 & 0.3 &-  & - & 1439.9/1350  \\
NGC3079  &0.9  & $201^{+28.5}_{-89.8}$ & $2.56^{+0.22}_{-0.23}$ & $0.84^{+0.02}_{-0.03}$& $700^{+200}_{-210}$ & 38.4 & 1.8  & 0.01  & 0.14 & 301.9/245 \\
NGC3147  &2.9  & $0.07^{+0.01}_{-0.01}$ & $1.56^{+0.04}_{-0.03}$ & - & $144^{+101}_{-55}$ & 155.8 & 31.1 & - & - & 562.1/585  \\
NGC3227 & 2.0  & $6.42^{+0.15}_{-0.17}$ & $1.44^{+0.04}_{-0.04}$ & -  & $152^{+22}_{-41}$ & 866.3  & 43.9 & 0.06 & 10.12 & 2089.4/2082  \\
NGC3254 & 1.8  & $<0.12$ & 1.9 (fixed) & -  & -  &  0.7  & 0.05  & -  & -  & 4.23/5  \\
NGC3486 & 1.7  & $8.42^{+28.9}_{-4.17}$ & 1.9  (fixed) & $0.37^{+0.06}_{-0.05}$  & -  &  8.5  &  0.06 & - & -  & 30.29/44 \\  
NGC3516$^{\star}$  & 3.5 & $1.04^{+0.32}_{-0.49}$ & $1.58^{+0.02}_{-0.04}$ &  -  & $180^{+13}_{-16}$  &  1460.3  & 264.1  & - &- & 2483.9/2056  \\
NGC3735  &1.3  & $15.23^{+55.43}_{-10.17}$ & $2.85^{+0.67}_{-0.49}$ & - & -  &  17.21  & 3.4  & 0.11  & -  & 139.7/131 \\ 
NGC3941  &1.9  & $<0.09$ & $2.02^{+0.35}_{-0.24}$ & - & -  &  4.19  & 0.2  &  &  & 87.97/84  \\
NGC3976  &1.1  & $0.12^{+0.07}_{-0.06}$ & $2.01^{+0.04}_{-0.03}$ & -  & -  &  7.3  & 1.2  & - & - & 109.42/109  \\
NGC3982  & 1.0 & $43.23^{+28.11}_{-16.96}$ & $2.53^{+0.44}_{-0.42}$ & $0.28^{+0.04}_{-0.04}$ & $802^{+678}_{-420}$  & 16.60 & 0.6  & 0.025  & - & 109.42/109 \\ 
NGC4051$^{\star}$  & 1.2 & $0.53^{+0.21}_{-0.35}$ & $2.08^{+0.11}_{-0.18}$ &   & $155^{+14}_{-18}$  & 650.32 & 22.4  & -  & 9.7 & 2315.34/2225  \\
NGC4138  & 1.3 & $8.85^{+0.41}_{-0.44}$ & $1.63^{+0.05}_{-0.05}$ & -  & $80^{+41}_{-34}$  &  554.3 & 19.1 & 0.009  & - & 473.04/449  \\
NGC4151$^{\star}$  & 2.3 & $5.14^{+0.13}_{-0.14}$ & $1.55^{+0.03}_{-0.03}$ & -  & $69^{+11}_{-10}$  & 21000.5 & 1034.4  & -  & 2.54  & 109.42/109 \\  
NGC4168  &2.4  & $<0.06$ & $2.02^{+0.14}_{-0.12}$ & -  & -  &  0.6  & 0.2  &  - & -  & 72.2/73 \\
NGC4169  &1.7  & $13.47^{+6.07}_{-3.31}$ & $2.01^{+0.79}_{-0.44}$ & -  &  & 23.5  & 6.9  & 0.017  &-  & 22.87/30 \\ 
NGC4258$^{\star}$  &1.6  & $6.8^{+0.35}_{-0.28}$ & $1.62^{+0.07}_{-0.05}$ & -  & $41^{+17}_{-19}$  &  380.9  & 2.1  &  -  & -  & 1227.17/1258  \\
NGC4378  &1.7  & $0.18^{+0.06}_{-0.06}$ & $1.55^{+0.19}_{-0.15}$ & -   &   &  14.7  & 2.1  &  &  & 31.84/38  \\
NGC4388  &2.6  & $32.14^{+1.19}_{-1.05}$ & $1.86^{+0.09}_{-0.08}$ & - & $173^{+15}_{-32}$ & 2007.3  &  70.8 & 0.006  &  & 1050.7/923 \\ 
NGC4395$^{\star}$  & 1.9 & $1.02^{+0.11}_{-0.10}$ & $1.18^{+0.03}_{-0.02}$ & -  & $80^{+16}_{-9}$  &  590.5  & 1.2  & - & - & 2298.85/2008  \\
NGC4472$^{\star}$  & 1.5 & $<0.82$ & $1.65^{+0.26}_{-0.14}$ & $0.87^{+0.15}_{-0.11}$ & -  &  21.5  &  0.7 & - & - & 283.25/292 \\  
NGC4477  &2.4 & $<0.02$ & $2.12^{+0.25}_{-0.17}$ &  $0.41^{+0.04}_{-0.16}$  & -  &  3.7  & 0.1  & - & - & 62.9/58  \\
NGC4501  &2.6  & $<0.06$ & $1.95^{+0.19}_{-0.18}$ &  $0.66^{+0.01}_{-0.01}$  & -  &  0.6  & 0.2  &-  &-  & 40.31/32  \\
NGC4565  &1.2  & $0.16^{+0.03}_{-0.03}$ & $1.87^{+0.14}_{-0.09}$ &  & -  &  20.7  & 0.2  &-  &-  & 102.5/88  \\
NGC4639  &2.2 & $<0.04$ & $1.79^{+0.06}_{-0.05}$ &  -  & -  &  48.2  & 1.6  & - &-  & 268.42/250 \\
NGC4698  &1.8 & $<0.07$ & $1.73^{+0.22}_{-0.27}$ &  -  & -  &  4.8  & 0.1  & - & - & 33.82/33  \\
NGC4725  &0.8  & $<0.03$ & $2.68^{+0.27}_{-0.23}$ & $0.23^{+0.03}_{-0.03}$ & -  &  2.3  & 0.04  & - & - & 64.4/71  \\
NGC5033 & 1.1  & $<0.04$ & $1.72^{+0.02}_{-0.02}$ & -  & $286^{+81}_{-71}$ & -  &  440.6  & 18.3  & - &  999.1/974  \\
NGC5194$^{\star}$  &1.8  & $<0.65$ & $1.16^{+0.14}_{-0.23}$ & -  & $1730^{+422}_{-275}$  &  25.5  &  0.2 & -  & -  &89.33/91  \\
NGC5273$^{\star}$  &0.9  & $0.72^{+0.07}_{-0.09}$ & $1.44^{+0.07}_{-0.09}$ & -  & $191^{+52}_{-72}$  &  706.7  & 38.2  &-  &-  & 612.24/635  \\
NGC6951  &12.4  & $0.40^{+0.31}_{-0.15}$ & $2.59^{+0.53}_{-0.30}$ & $0.67^{+0.16}_{-0.18}$  & -  &  4.8  & 0.3  & - & - & 32.33/55 \\ 
NGC7479  &5.3  & $40.2^{+11.44}_{-8.75}$ & $2.56^{+0.15}_{-0.16}$ & $0.27^{+0.07}_{-0.07}$ & $480^{+210}_{-390}$  &  22.5  & 2.8 & 0.004 & - & 137.3/133 \\ 
NGC7743  &4.8  & $0.3^{+0.18}_{-0.14}$ & $3.63^{+1.14}_{-1.03}$ &  $0.23^{+0.06}_{-0.06}$ & -  &  1.0  & 0.07  & - & -  & 16.7/18  \\
\end{tabular}                                                 
\begin{list}{}{}
\item{Column 1: Galaxy name}
\item{Column 2: Galactic column density in units of $10^{20}$ cm$^{-2}$}
\item{Column 3: Observed column density in units of $10^{22}$ cm$^{-2}$}
\item{Column 4: Power-law photon index of the continuum emission}
\item{Column 5: Temperature of the Raymond-Smith model}
\item{Column 6: Equivalent width of the FeK$_{\alpha}$ emission}
\item{Column 7: Observed 2-10 keV flux in units of $10^{-14}$ ergs s$^{-1}$ cm$^{-2}$}
\item{Column 8: Observed 2-10 keV luminosity in units of $10^{40}$ ergs s$^{-1}$}
\item{Column 9: Ratio of the normalizations of the scattered to the continuum emission}
\item{Column 10: Ratio of the normalizations of the reflected to the continuum emission}
\item{Column 11: Reduced $\chi^2$}
\item{$^\star$ Indicates source with very complex spectra for 
which only a rough parametrization in the 2-10 keV band is presented here.}
\end{list}    
\end{table*}

The \xmm X-ray spectra of our sources are presented in 
Fig \ref{xspectra}. For each object the upper panel shows the X-ray spectrum 
along with the model presented in Table \ref{fit} while the lower panel shows the residuals.

\section{X-ray absorption}

The spectral fitting results are presented in Table \ref{fit}.
There are 8 type-1 Seyferts in our sample.  
Five of them show small amounts of absorption  
($<10^{21}$ cm$^{-2}$) while the 3  Seyfert-1.5 sources 
(NGC3227, NGC3516, and NGC4151) present  a considerable amount of 
$N_H$ ($>10^{22}$ cm$^{-2}$).
Our sample contains 30 Seyfert-2 galaxies. 
The column densities in this population vary from the Galactic 
to the Compton-thick limit ($N_H>$10$^{24}$ cm$^{-2}$).
However, the apparent number of significantly obscured sources 
is rather small. Only 12 out of 30 type-2 sources present 
absorption greater than $10^{22}$ cm$^{-2}$.

\subsection{Compton-thick sources}

The fraction of Compton-thick sources 
 is more difficult  to estimate.   This is because 
the \xmm  effective area   
sharply decreases at energies higher than 6 keV. 
Given the limited \xmm bandpass, which extends up to about 10 keV,
we are not able to measure the absorption turnover for highly 
absorbed sources. A column density of $\sim$10$^{24}$ cm$^{-2}$ 
suppresses  90 \% of the  flux in the 2-10 keV band. 
Therefore, we can obtain a direct measurement of the obscuration  
only up to column densities reaching at most a few times $10^{24}$ \cunits. 
In the case of Compton-thick AGN the X-ray spectrum is dominated by 
scattered components from cold or warm material as well as
an FeK$_\alpha$ with high equivalent width (Matt et al. 2000).
Then to  unveil the presence of a Compton-thick nucleus we apply  the 
following diagnostics.

\begin{itemize}
\item{ Flat X-ray spectrum ($\Gamma<1$). This  
implies the presence of a strong reflection component, 
which intrinsically flattens the X-ray spectrum at higher 
energies (e.g. Matt et al. 2000)}
\item{ High Equivalent Width of the FeK$_{\alpha}$ line ($\sim$1 keV). This 
characteristic is consistent with a  Compton-thick nucleus since 
then the line is  measured against a much depressed continuum  
(Leahy \& Creighton 1993) or a pure reflected component.}
\item Low X-ray to optical flux ratio. Bassani et al. (1999) 
have showed that the 2-10 KeV to the [OIII] $\lambda$5007 flux ratio is 
very effective in the identification of Compton-thick sources. 
This is because the [OIII] $\lambda$5007 (hereafter [OIII]) flux 
which comes from large (usually kpc) scales, remains unabsorbed 
while the X-ray flux is diminished because of absorption.
\end{itemize}

\begin{figure}
\centering
\rotatebox{0}{\includegraphics[width=8cm, height=8cm]{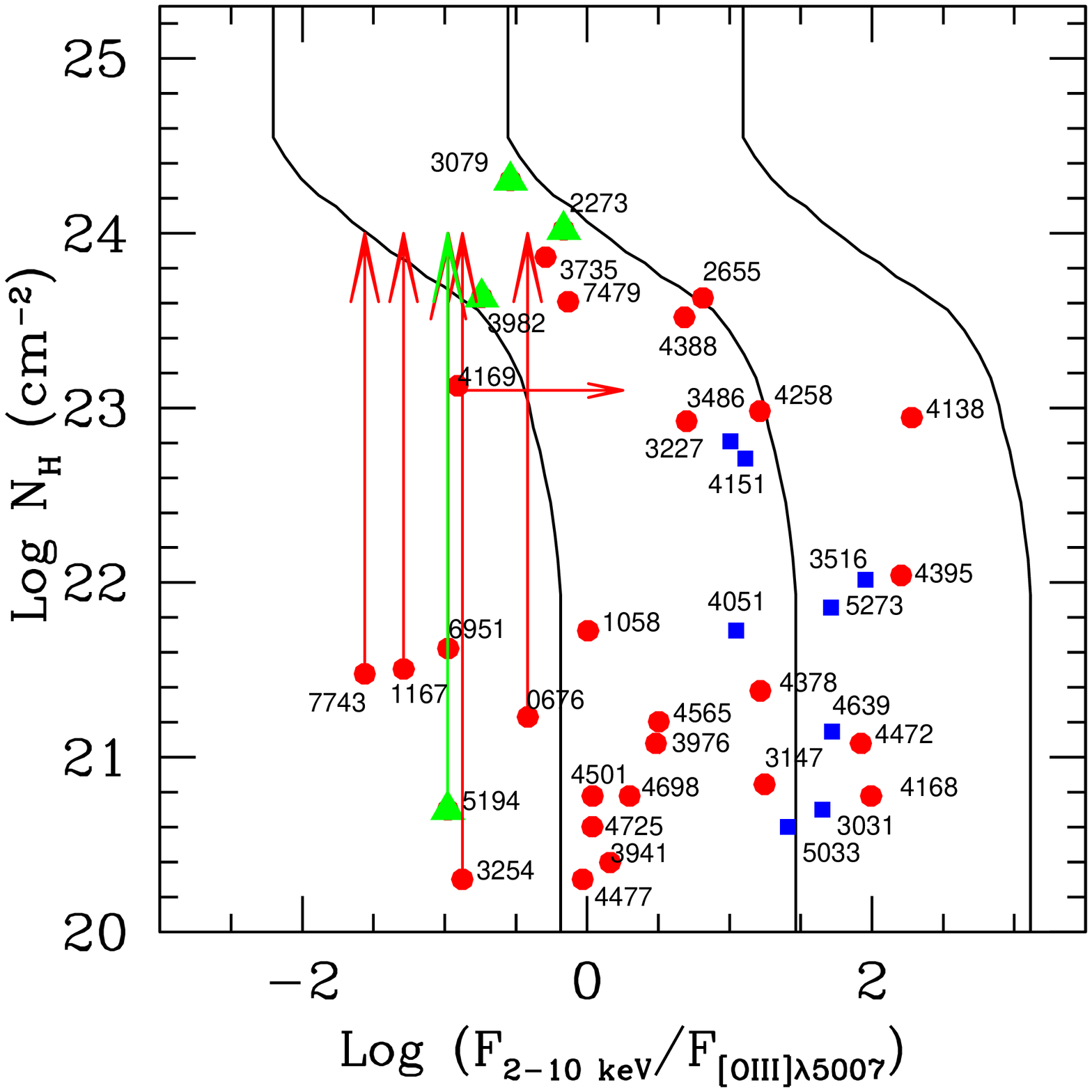}}
\caption{Distribution of the $N_H$ values as a function of the 
$F_{2-10~\rm keV}/F_{[OIII]}$ ratio. Filled boxes and circles denote 
Seyfert-1 and Seyfert-2 galaxies respectively. The triangles show 
Seyfert-2 galaxies with large EW($>$700 eV). The solid lines represent 
the mean $\rm N_H vs. F_{2-10 \rm keV}/F_{[OIII]}$ relation followed by the 
Seyfert-1 in our sample together with the $\pm3\sigma$ dispersion 
(see text).}
\label{oiii}
\end{figure}

These criteria however should be considered with caution. 
For example high Equivalent Width (EW) lines may also appear in the case 
of anisotropic distribution of the scattering medium 
(Ghisellini et al. 1991), or in the case where there is a time lag 
between the reprocessed and the direct component (e.g. NGC2992, Weaver et al. 1996). 
Also there have been reports of Compton-thick sources where the 
value of FeK$_{\alpha}$ line EW is well below 1 keV 
(e.g. Awaki et al. 2000 for Mkn1210). 

In Fig. \ref{oiii}  we plot the column density obtained from the spectral fittings 
as a function of the X-ray to optical flux ratio, 
$F_{2-10~\rm keV}/F_{[OIII]}$. The [OIII] fluxes are corrected for the optical reddening 
using the  formula described in Basanni et al. (1999): $F_{[OIII]_{\rm COR}}=
F_{[OIII]_{\rm OBS}}[(H_{\alpha}/H_{\beta})/(H_{\alpha}/H_{\beta})_{o}]^{2.94}$, 
where the intrinsic Balmer decrement $(H_{\alpha}/H_{\beta})_{o}$ equals 3.  

The solid lines in Fig. \ref{oiii} show the expected correlation 
between these quantities, assuming a photon index 
of 1.8 and  $1\%$ reflected radiation 
(see also Maiolino et al 1998, Cappi et al 2006). 
The starting point in the x-axis for the middle solid line is 
determined by averaging the $F_{2-10~\rm keV}/F_{[OIII]}$ values of 
the Seyfert-1 population only, while the lines at right and left 
show the $3\sigma$ dispersion. The sources occupying the low 
($F_{2-10~\rm keV}/F_{[OIII]}$, $ N_H$) region in this plot could be 
possibly highly obscured or Compton-thick AGN.

In two cases (NGC2273, NGC3079) we can immediately tell 
the presence of a Compton-thick  nucleus through the presence of 
an absorption turnover in the spectral fittings. 
Both sources present high values of the FeK$_{\alpha}$  
line EW  ($>700$ eV). 
One more source (NGC5194), despite the fact that it presents the 
highest value of FeK$_{\alpha}$ ($\sim 1700$ eV),    
shows no absorption at all.
However, the very flat X-ray spectrum and the very low 
$F_{2-10~\rm keV}/F_{[OIII]}$ value further suggest that this is a 
highly obscured or a Compton-thick source.
According to the $N_H$-$F_{2-10~\rm keV}/F_{[OIII]}$  relation
a minimum  value for the $N_H$ is $4\times10^{23}$ cm$^{-2}$
(see Fig. \ref{oiii}. 

\begin{figure}
\centering
\rotatebox{-90}{\includegraphics[width=6cm, height=8cm]{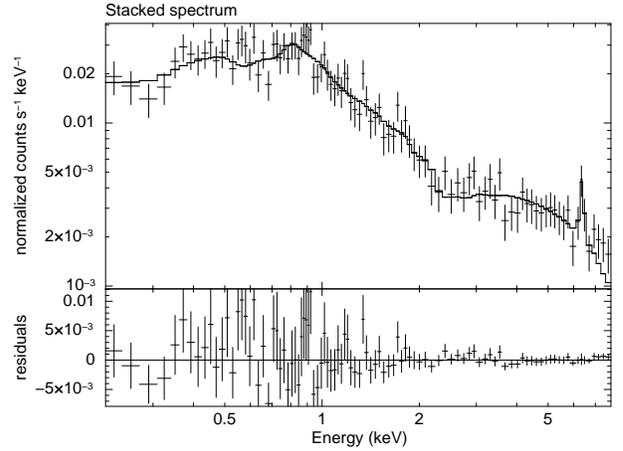}}
\caption{Stacked X-ray spectrum of the Seyfert-2 Galaxies NGC0676, 
NGC1167, NGC3254, NGC6951 and NGC7743. 
The best fit model and residuals are also shown.}
\label{stacked}
\end{figure}

There are also 5 Seyfert-2 galaxies (NGC676, NGC1167, NGC3254, 
NGC6951 and NGC7743) occupying the low $F_{2-10~\rm keV}/F_{[OIII]}$
regime. We do not consider  NGC4169 because of the large 
error in the estimation of the [OIII] flux (see Ho et al. 1997). 
These, according to the expected $N_H$-$F_{2-10~\rm keV}/F_{[OIII]}$ 
relation, should present high values of $N_H$. 
According to  Fig. \ref{oiii}, 
 the minimum $N_H$ value is $\sim2\times10^{23}$ cm$^{-2}$ 
for  NGC676, $\sim6\times10^{23}$ cm$^{-2}$ for NGC6951 and $\sim10^{24}$ 
cm$^{-2}$ for NGC1167 and NGC7743.
However, the X-ray spectral fittings show low absorption 
($<10^{22}$ cm$^{-2}$) while in addition there is no indication for a flat 
photon index or strong FeK$_\alpha$ line. 
This may be due to the limited photon statistics in the hard ($>2$ keV) band, 
which does not allow us to examine in detail the spectral characteristics.
 Note however, that in the spectrum of at least two sources  (NGC1167 and NGC7743)
there is some indication for a flattening at hard energies 
 which could suggest a heavily buried or reflected component. 
We investigate further this issue by deriving the mean, stacked X-ray spectrum. 
 We use the {\sl MATHPHA} tasks of {\sl FTOOLS} 
to derive the weighted stacked X-ray spectrum of the five 
EPIC-PN observations. The corresponding  ancillary files are 
also produced using {\sl  ADDRMF} and {\sl ADDARF} tasks of {\sl FTOOLS}.
We perform no correction for the rest-frame energy 
because the differences in the redshifts are  negligible.
An absorbed power-law model plus a Gaussian line  
and a soft excess component (Raymond-Smith model) 
reproduce well the mean spectrum (Table \ref{stack_table}).
In Fig. \ref{stacked} we present the data along with the best-fit.
The average spectrum shows significant absorption  consistent with the measured value 
of the FeK$_{\alpha}$ line EW.

\begin{table*}
\centering
\caption{The stacked X-ray spectrum of the 5 Seyfert-2 galaxies with low $F_{2-10~\rm keV}/F_{[OIII]}$}
\label{stack_table}
\begin{tabular}{cccccccccc}
$N_H$ (cm$^{-2}$) & {$\Gamma$} & kT (keV) & EW$_{\rm FeK}$ (eV) & $\chi^2_{\nu}$ \\
\hline
\hline
$10.30^{+2.01}_{-1.91}$ & $1.71^{+0.08}_{-0.17}$ & $0.73^{+0.13}_{-0.12}$ & $255^{+243}_{-146}$ & 72.41/91  \\
\hline
\end{tabular}
\end{table*}

\begin{figure}
\centering
\rotatebox{0}{\includegraphics[width=8cm, height=6cm]{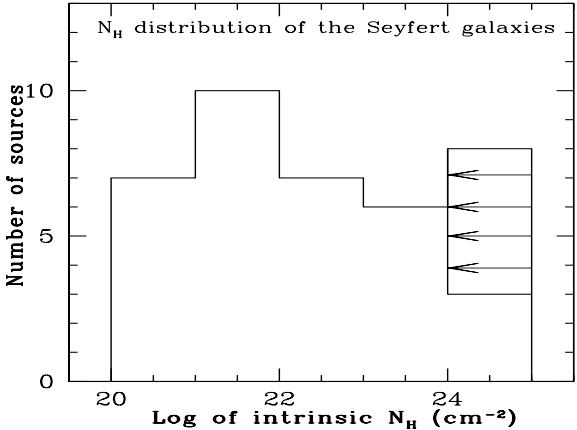}}
\caption{The estimated $N_H$ distribution of the Seyfert galaxies 
in our sample (solid line). The vertical arrows in the last $N_H$ 
bin (Log$(N_H)>$24 cm$^{-2}$) show the upper and the lower 
limits for the number of the the Compton-thick sources in our sample.}
\label{nh}
\end{figure}

Our results above can be summarised as follows. 
The number of absorbed nuclei ($N_H>10^{22}$ cm$^{-2}$) are 21 out of 38 
or $55\pm12$\%. The number of Compton-thick sources 
is three i.e. $8\pm5$\% although, if we adopt the extreme case 
where all the low   $F_{2-10~\rm keV}/F_{[OIII]}$  host Compton-thick 
nuclei this  number would rise to  8 or $21\pm 7$ \%. 
Our estimates on the amount of $N_H$ in the local universe 
are illustrated in Fig. \ref{nh}. 
The solid line describes the $N_H$ distribution. 
The vertical arrows in the highest $N_H$ bin show the 
upper and lower limits for the number of 
Compton-thick sources.

\subsection{The absorption in the bright sub-sample} 

\begin{figure}
\centering
\rotatebox{0}{\includegraphics[width=8cm, height=6cm]{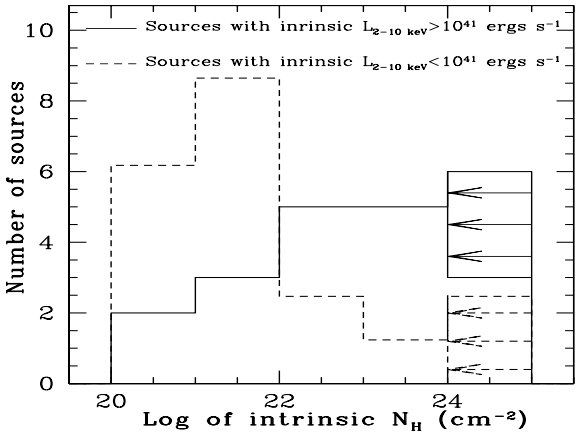}}
\caption{The derived $N_H$ distribution of the Seyfert galaxies 
in the bright (solid line) and the faint (dashed line) sub-samples. 
The vertical arrows in the Compton-thick regime 
(Log$(N_H)>$24 cm$^{-2}$) define the upper and lower limit for 
the number of the Compton-thick sources in our sample.}
\label{nh_lum}
\end{figure}

Our findings should play an important role to the XRB synthesis models
(Comastri et al. 1995, Gilli et al. 2007). Gilli et al. 2007 assume
in their models a lower luminosity  of  10$^{41}$ ergs s$^{-1}$.
However, the  intrinsic 2-10 keV luminosity  of our sources 
starts from as low as a few times 10$^{38}$ ergs s$^{-1}$  which is about 
3 orders of magnitude lower. Therefore it is 
useful to present our results separately for the fainter 
($L_{2-10~\rm keV}<10^{41}$ ergs s$^{-1}$) and the brighter 
($L_{2-10~\rm keV}>10^{41}$ ergs s$^{-1}$) sub-sample containing 
21 and 17 sources respectively (see Fig. \ref{nh_lum}). 
The intrinsic $L_X$ values are 
determined using the best fitting results. 
For the three Compton-thick sources the intrinsic $L_X$ 
has been estimated assuming that 1$\%$ of the intrinsic luminosity is 
actually observed below 10 keV due to scattering and or reflection 
(e.g. Comastri 2004). 
In the bright sample the fraction of the highly absorbed 
sources is  $\sim$75 \% and the Compton-thick sources
most probably account for 15-20 $\%$ of the total population.
This fraction can reach a maximum of 29 $\%$ in the unlikely 
case where all the low $F_{2-10~\rm keV}/F_{[OIII]}$ sources  
host a Compton-thick nucleus.

\subsection{Unabsorbed Seyfert-2 Galaxies}

The X-ray spectral analysis reveals several 
Seyfert-2 galaxies with very little or no X-ray absorption. 
As we have already discussed some of these, 
i.e. the five with low X-ray to [OIII] flux ratio  
are most probably associated with a highly obscured or 
even a Compton-thick nucleus.
In Fig. \ref{oiii} there are 12 additional Seyfert-2 galaxies 
(NGC1058, NGC3147, NGCC3941, NGC3976, NGC4168, NGC4378, NGC4472, NGC4477, 
NGC4501,NGC4565,NGC4698 and NGC4725) with $N_H$ less than 
10$^{22}$ cm$^{-2}$ but an average  $F_{2-10~\rm keV}/F_{[OIII]}$ value. 
This behaviour is not unknown (e.g. Pappa et al 2001, 
Gliozzi, Sambruna \&  Foschini  2007). In particular NGC3147 is a well 
established example, through simultaneous optical and X-ray observations, of
a spectroscopically classified Seyfert-2 galaxy 
with very little or no absorption (Bianchi et al. 2008).
NGC4698 and NGC4565 have also been discussed to be good candidates, 
(see Georgantopoulos \& Zezas 2003, Panessa \& Bassani 2002).

It is possible that some of our new unabsorbed candidates are 
contaminated by nearby luminous X-ray sources that we are unable 
to resolve owing to the X-ray telescope's angular resolution. 
An inspection  of the available \chandra images which 
have a superior resolution (0.5 arcsec FWHM)   
could be very helpful towards this direction. 
All but three sources (NGC3941, NGC3976 and NGC4378) present archival 
\chandra data.
Although a detailed  analysis of the properties of the unabsorbed 
Seyfert-2 galaxies is the scope of a forthcoming paper,
we briefly report on whether  there is any evidence for contamination.
NGC1058 and NGC4168 are significantly contaminated from nearby 
luminous X-ray sources (see also Foschini et al. 2002, Cappi et al 2006) 
while NGC4472 suffers from very strong diffuse emission. 
Finally, inspection of \xmm images show that 
NGC3941 and NGC4501 are contaminated (less than 30\% of the counts) 
by nearby sources (see also Foschini et al. 2002, Cappi et al 2006). 

We further try to examine the X-ray  properties of unobscured 
Seyfert-2 galaxies by deriving their stacked spectrum. 
We use {\sl MATHPHA} task of {\sl FTOOLS} software to create the 
weighted mean X-ray spectrum of the EPIC-PN observations. 
Weighted mean ancillary files are produced using the {\sl  ADDRMF} and 
{\sl ADDARF} tasks of {\sl FTOOLS}.
NGC3147, NGC4565 and NGC4698 are not considered in the mean spectrum 
since there is already evidence that they do not present any 
absorption. We also exclude the five contaminated sources leaving
the cases of NGC3976, NGC4725, NGC4378 and NGC4477 to be considered.

\begin{figure}
\centering
\rotatebox{-90}{\includegraphics[width=6cm, height=8cm]{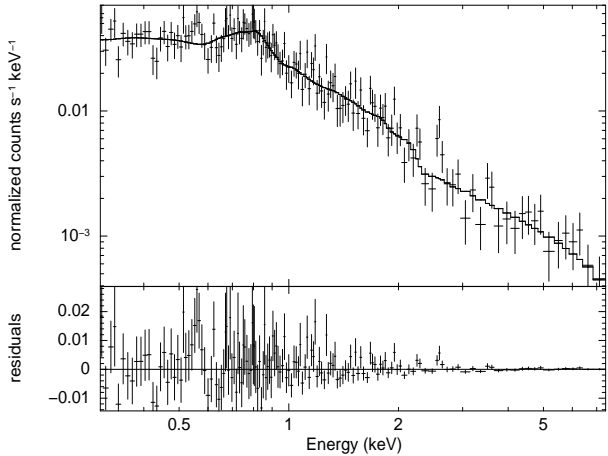}}
\caption{Stacked X-ray spectrum of the unabsorbed Seyfert-2 Galaxies.
The best fit model and the residuals are also shown.}
\label{stacked1}
\end{figure}

We try to detect any spectral feature, such as the FeK$_{\alpha}$ 
line, that could give away the presence of a hidden nucleus in this 
population as marginally suggested in some cases 
(e.g. Brightman \& Nandra 2008). We fit the average spectrum with 
an absorbed power-law model plus a Raymond-Smith model. 
The spectral fitting results are listed in table \ref{unabsorbed}. 
There is no significant evidence for the presence of an FeK$_\alpha$ 
emission line.  Nevertheless, if we choose to include a Gaussian 
component, the upper limit of the EW is $\sim$600 eV at the 90 per cent 
confidence level. In Fig. \ref{stacked1} we present the mean 
spectrum along with the best fit model and the residuals. 
Assuming that all these sources are truly unabsorbed Seyfert-2 galaxies
then their total fraction accounts for $\sim$20 per cent of the total population.

\begin{table*}
\centering
\caption{The stacked X-ray spectrum of the unabsorbed Seyfert-2 galaxies}
\label{unabsorbed}
\begin{tabular}{cccccccccc}
$N_H$ (cm$^{-2}$) & {$\Gamma$} & $\rm kT$ (keV) & EW$_{\rm FeK}$ (eV) & $\chi^2_{\nu}$ \\
\hline
\hline
$<0.1$ & $2.02^{+0.18}_{-0.15}$ & $0.37^{+0.07}_{-0.08}$ & $<600$ & 118.6/125  \\
\hline
\end{tabular}
\end{table*}

It has been proposed that the unabsorbed Seyfert-2 
galaxies are 'naked' nuclei i.e. they lack a 
Broad-Line-Region, BLR, (see Ho 2008 for a review). 
Various theoretical models could explain this behaviour.     
Nicastro (2000) presented a model which relates the width of the 
Broad Emission Lines of AGN to the Keplerian velocity of an 
accretion disk at a critical distance from the central black hole. 
Under this scheme the Broad Line Region is linked to the 
accretion rate of the AGN i.e. below a minimum accretion rate the BLR cannot form.
Recently Elitzur \& Shlosman (2006) presented an alternative model which 
depicts the torus as the inner region of clumpy wind outflowing 
from the accretion disc. According to this  model the 
torus and the BLR disappear when the bolometric luminosity decreases 
below $\sim 10^{42}$ ergs s$^{-1}$ because the accretion onto the central 
black hole can no longer sustain the required cloud outflow rate.

In Table \ref{model} we try to compare our results with the 
above model predictions. 
In Col. 1 we give an estimate for the mass of the central 
Black Hole, taken from Panessa et al. (2006) and McElroy (1995).
The mass estimation is inferred from the mass-velocity dispersion 
correlation. In the case of NGC3976 there is no information available in 
the literature. In Col. 2 we calculate the bolometric 
luminosities using the corrections determined by Elvis et al. (1994) i.e., 
$L_{\rm BOL}$ = 35 $\times L_{2-10~\rm keV}$ ergs s$^{-1}$. 
In Col. 3 we give the accretion rate estimator given by 
$L_{\rm BOL}$/$L_{\rm EDD}\simeq1.3 \times 10^{38} M/M_{\odot}$. 
The Eddington Luminosity ($L_{\rm EDD}$) is given by 
$L_{\rm EDD}$=$4\pi G M m_p c / \sigma_T$ where M is the black 
hole mass, $m_p$ is the proton mass, $\sigma_T$ is the 
Thomson scattering cross section.

All the sources present  very low accretion rates, well 
below the threshold of 1-4 $\times10^{-3}$ proposed by Nicastro 
(2000) and Nicastro et al. (2003).
Furthermore all these sources (but NGC3147) present very low bolometric 
luminosities also below the critical value of $10^{42}$ \lunits predicted by 
Elitzur \& Shlosman (2006). This supports the idea that the key 
parameter is not the orientation but an intrinsic parameter 
(low accretion rate or luminosity), which prevents the formation of the BLR.

\begin{table*}
\centering
\caption{Accretion rates and luminosities for the unabsorbed Seyfert-2 galaxies}
\label{model}
\begin{tabular}{cccccc}
Name & Log($M_{BH}/ M_{\odot})$ & $Log(L_{\rm BOL})$ & $L_{\rm BOL}$ / $ L_{\rm EDD}$  \\
     &                          &                &       $\times 10^{-4}$     \\                   
\hline
NGC 1058 & 4.9 & 39.8 &  5.5 \\
NGC 3147 & 8.8 & 43.0 &  1.1 \\
NGC 4168 & 7.9 & 41.3 &  0.15 \\
NGC 4378 & 7.9 & 41.8 &  0.50 \\
NGC 4472 & 8.8 & 41.4 &  0.026 \\
NGC 4477 & 7.9 & 40.5 &  0.025 \\
NGC 4501 & 7.9 & 41.3 &  0.17 \\
NGC 4725 & 7.5 & 40.1 &  0.027 \\
NGC 4565 & 7.7 & 40.8 &  0.084 \\
NGC 4698 & 7.8 & 40.5 &  0.030 \\
\end{tabular}
\begin{list}{}{}
\item{Column 1: Name}
\item{Column 2: Black Hole mass in units of Solar Masses}
\item{Column 3: Bolometric luminosity in units of ergs s$^{-1}$}
\item{Column 4: Accretion rate  $L_{\rm BOL}/L_{\rm EDD}$}
\end{list}{}{}

\end{table*}

\section{Discussion}

\subsection{Comparison with other optically selected samples}
 
In this work we present \xmm  observations of  all the Seyfert galaxies 
from the Palomar survey (Ho et al. 1995).  
We find that  $\sim$50 per cent of the Seyfert population 
is absorbed by $N_H>10^{22}$ cm$^{-2}$. 
In this sample we have identified 3 Compton-thick sources
which translates to a fraction of $\sim$8 per cent.
Five more sources possibly host a highly absorbed or 
a Compton thick nucleus. In the very 
extreme, and rather unlikely case were all these  candidates are true  
Compton-thick sources  their fraction reaches 20 per cent of the 
total population.

Cappi et al. (2006) and Pannesa et al. (2006), also using data 
from the Palomar survey, provide estimates for the fraction 
of obscured AGN in the local universe. These authors find that 
about 50\% of their sources are obscured ($>10^{22}$ \cunits). 
Their  estimates on  the fraction of Compton thick sources suggest 
an absolute minimum of 20 per cent of the total population. 
This result comes in contradiction with our findings.
However their sample includes 2 objects not fulfilling the  
Palomar Survey selection criteria (see also Section \ref{thesample}). 
These are the 2 Compton-thick AGN NGC1068 and NGC3185.
When we exclude these an agreement is found.

Risaliti et al. (1999) study the X-ray absorption in a sample of 
45 Seyfert-2 galaxies finding that a  considerable fraction 
of these  are associated with Compton-thick nuclei.
A direct comparison with our results is not 
straightforward since these authors exclude all the sources with  
$F_{[OIII]}>4\times 10^{-13}$ \funits. 
However we  think that only a luminosity cutoff could 
reveal column density distribution of the population that 
contributes to the XRB (see Section 5.2).

\subsection{X-ray background synthesis models} 
 
The XRB synthesis models can provide 
tight constraints on the number density of Compton-thick sources.
These models attempt to fit the spectrum of the X-ray background 
roughly in the 1-100 keV range. 
It is well established that a large number of Compton-thick 
sources is needed (Gilli et al. 2007) to reproduce the hump 
of the X-ray background spectrum at 30-40 keV where most of its energy 
density lies (Churazov et al. 2007, Frontera et al. 2007). 
Here, we compare the fraction of the Compton-thick sources 
predicted by the model of Gilli et al. 2007 with our  results.
We use the publicly available POMPA software \footnote{www.bo.astro.it/$\sim$gilli/counts.html}.
This predicts the number counts at a given redshift, 
flux and luminosity range  using the best-fit results for the fraction of 
obscured, Compton-thick sources, of the X-ray background synthesis 
model of Gilli et al. (2007).

We restrict the comparison to low redshifts $z<0.017$ and 
X-ray luminosities in the bin $\rm 10^{41}<L_{2-10~\rm keV}<10^{44}$.
As our sample is not flux limited in the X-rays,   
we have to choose a flux limit deep enough to ensure that all 
sources in this luminosity and redshift bin are detected.
A flux limit of $10^{-14}$ {\funits} satisfies this constraint.  
The X-ray background synthesis models predict a fraction of Compton-thick 
sources  of about 40 \% which is higher compared with our results $15\pm8$\%.
Only if all the low $F_X/F_{[OIII]}$ Seyfert-2 galaxies are 
associated with a Compton-thick nuclei the discrepancy would become 
less pronounced.

Optically selected samples can still miss a fraction of
Compton-thick AGN. For example, NGC6240 is classified as a LINER  
in the optical while {\it BeppoSAX} observations show the presence 
of a Compton-thick nucleus (Vignati et al. 1999). Moreover 
{\it SUZAKU} observations (Ueda et al. 2007 and Comastri et al. 2007) 
have demonstrated that a small fraction of AGN may have a 4$\pi$
coverage, instead of the usually assumed toroidal structure. 
These sources will not exhibit the usual high excitation narrow 
emission lines and therefore will not be classified as AGN on the 
basis of their optical spectrum.  
 
Recent results  based on {\it INTEGRAL} and  {\it SWIFT} 
observations reveal a small fraction of Compton-thick 
sources (e.g. Sazonov et al. 2008, Sazonov et al. 2007, 
Ajello et al. 2008).
In particular at the flux limit of $\sim 10^{-11}$ \funits
in the 17-60 keV energy band, \integral observations find 
$10-15\%$ Compton-thick sources. The {\it SWIFT}/BAT hard 
X-ray survey failed to identify any Compton-thick AGN. 
This non detection discards the 
hypothesis that their fraction accounts for the 20 per cent of 
the total AGN at $>2\sigma$ confidence level. It is true however 
that some heavily obscured Compton-thick sources with 
$N_H\sim 10^{25-26}$ \cunits would be missed even by 
these ultra hard X-ray surveys.

\subsection{Less absorption at very low luminosities}  
 In the  low-luminosity sub-sample (intrinsic $L_{2-10~\rm keV}<10^{41}$ 
erg s$^{-1}$) the fraction of obscured sources  
diminishes to 30\%. This result comes in apparent contradiction  
with recent findings suggesting an increasing fraction of obscuration 
with decreasing luminosity 
(e.g.  Akylas et al. 2006, La Franca et al. 2005).
This behaviour may reflect a physical dependence of the column density 
with intrinsic luminosity as suggested by Elitzur \& Shlosman (2006). 
These authors present a model where the torus and the BLR disappear when the 
bolometric luminosity decreases below $\sim 10^{42}$ ergs s$^{-1}$ 
because the accretion onto the central black hole can no longer 
sustain the required cloud outflow rate. 
It is interesting to note that the corresponding luminosity
 in the 2-10 keV band is about  several $\times 10^{40}$ \lunits,
 assuming the Spectral Energy Distribution of Elvis et al. (1994). 
 Interestingly, almost all of our Seyfert-2 sources with no absorption 
 present luminosities below this limit 
(with the exception of NGC3147). We note however,  
there are sources (NGC3486, NGC3982) with low luminosity, which present column 
densities around $10^{22}$-$10^{23}$ cm$^{-2}$. 
 Alternatively, it is possible that at least in a few cases, 
 the large XMM-Newton Point Spread Function results in contamination 
 by nearby sources. Thus the nuclear 
 X-ray emission could be out-shined giving the impression that there 
 is no obscuration (e.g. Brightman \& Nandra 2008). 
 However, both the inspection 
 of the \chandra images as well as the stacked spectrum   
 of the unabsorbed  sources do not favour such a scenario. 

\section{Conclusions}
\xmm observations are available for all 38 Seyfert galaxies 
from the Palomar spectroscopic sample of galaxies of 
Ho et al. (1995, 1997).  Our goal is to determine the 
distribution of the X-ray absorption in the local Universe
through X-ray spectroscopy. 
Our sample consists of 30 Seyfert-2 and 8 Seyfert-1  galaxies. 
The results can be summarised as follows:  
 
\begin{itemize} 
\item{We find a high fraction  of obscured sources ($>10^{22}$ \cunits)  of about 50 \%. }

\item{A number of sources present low $F_X/F_{[OIII]}$ ratio. 
Their individual spectra show no evidence of high absorbing column densities. 
However, their stacked spectrum shows significant amount of absorption ($\sim 3\times 10^{23} $ \cunits)}  
 
\item{ Considering only the bright sub-sample ($L_{2-10~\rm keV}>10^{41}$ \lunits) ,
i.e. only these sources which contribute a significant amount to the X-ray 
background flux, we find that 75 \% of our sources are obscured.}
   
\item{ In the bright sub-sample there are at least 3 Compton-thick AGN translating to a fraction of 
15\% which is lower than the predictions of the X-ray background synthesis models at this luminosity and 
redshift range. Only if we consider, the rather unlikely scenario, where all Seyfert-2 galaxies 
with a low $F_X/F_{[OIII]}$ ratio are associated with Compton-thick sources we would alleviate 
this discrepancy.}  
   
\item{ We find a large number of unobscured Seyfert-2 galaxies.   All these have low luminosities 
$L_{2-10~\rm keV } <3\times 10^{41}$ \lunits.  Inspection of the \chandra images, where available, 
demonstrates that in most cases these are not contaminated by nearby sources. 
Furthermore, their stacked spectrum reveals no absorption. It is most likely that 
these are genuinely unobscured sources in accordance with the predictions 
of the models of Elitzur \& Sloshman (2006)}.  
\end{itemize}

\begin{figure*}
\centering
\rotatebox{-90}{\includegraphics[width=6cm, height=8cm]{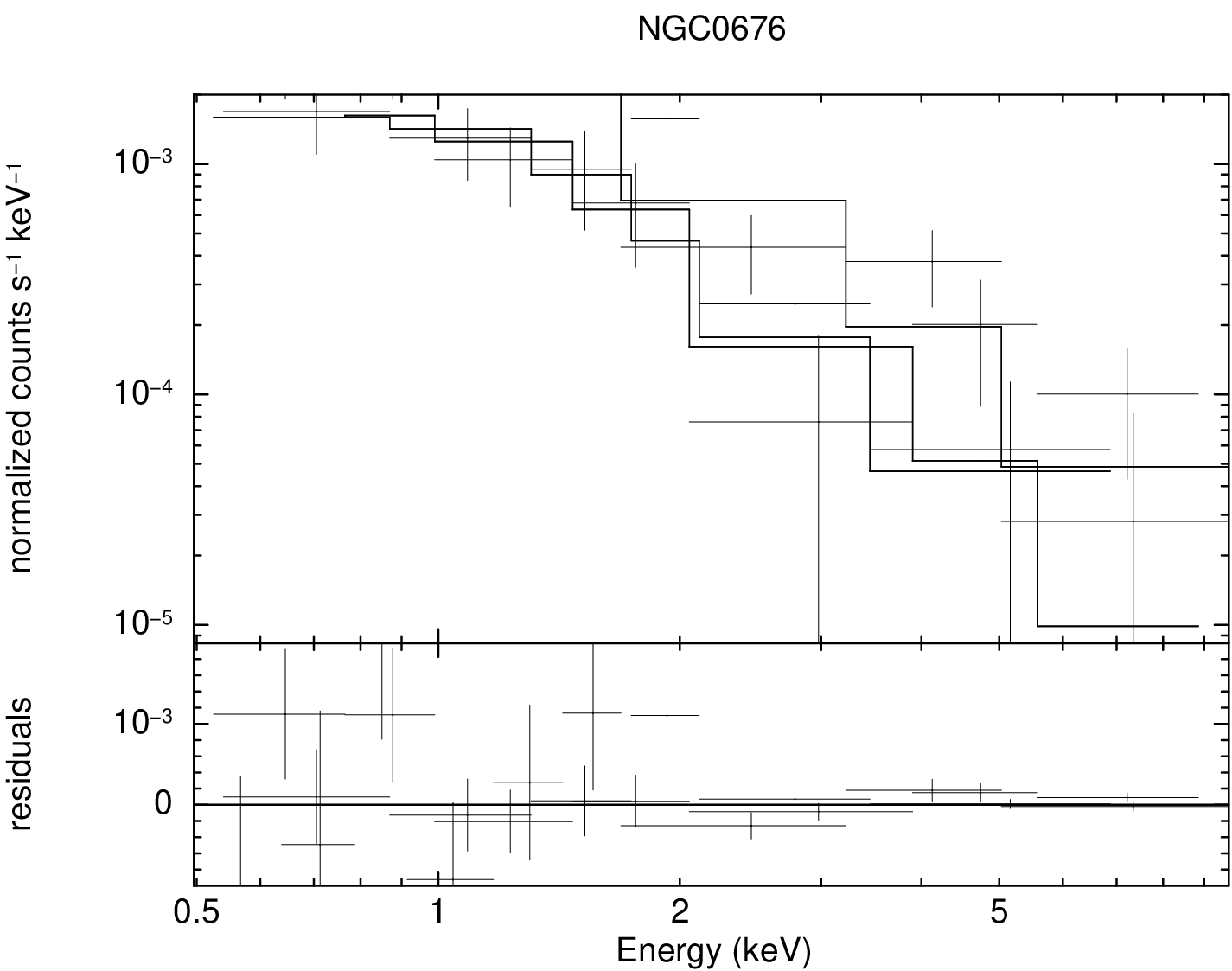}}
\rotatebox{-90}{\includegraphics[width=6cm, height=8cm]{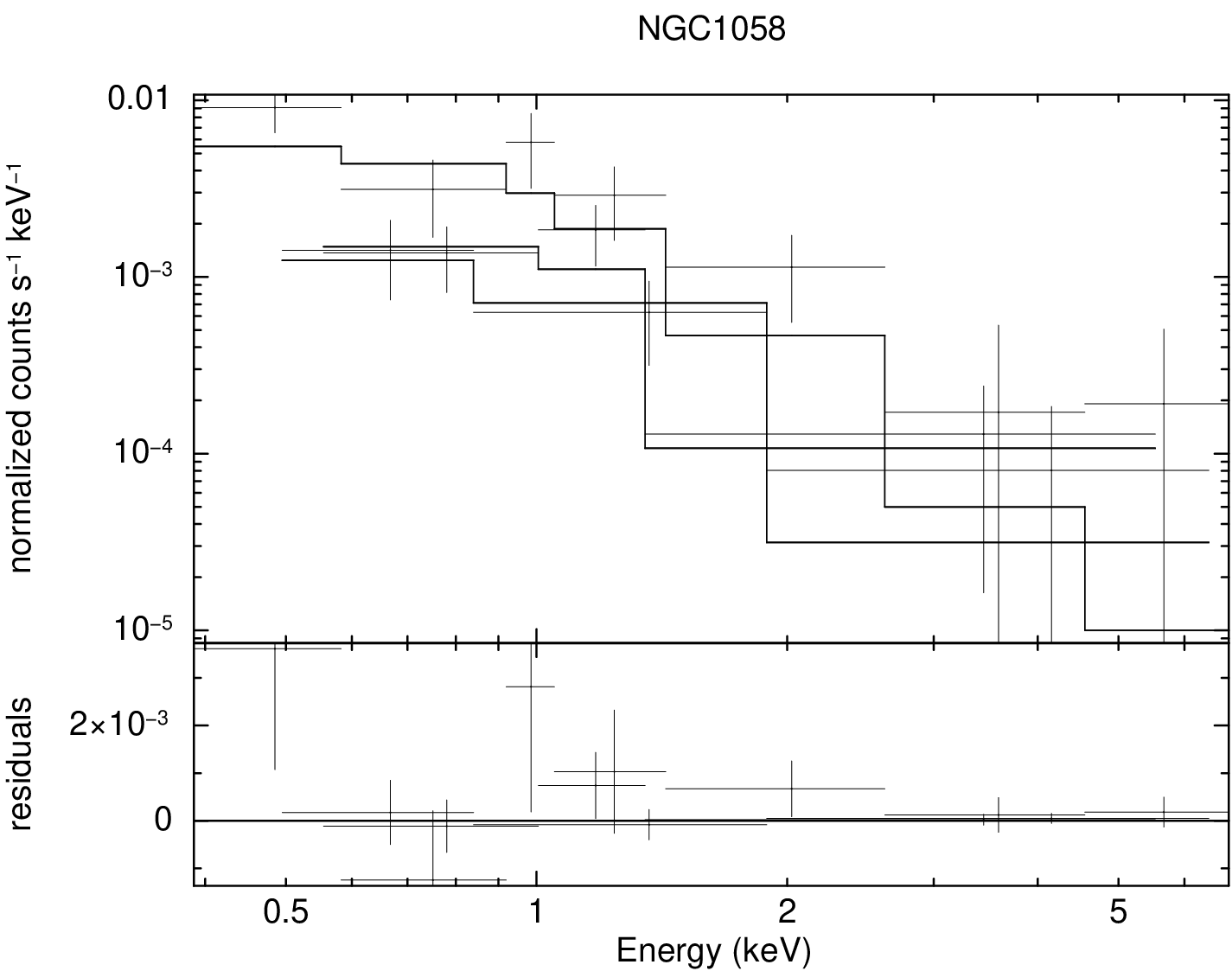}}
\rotatebox{-90}{\includegraphics[width=6cm, height=8cm]{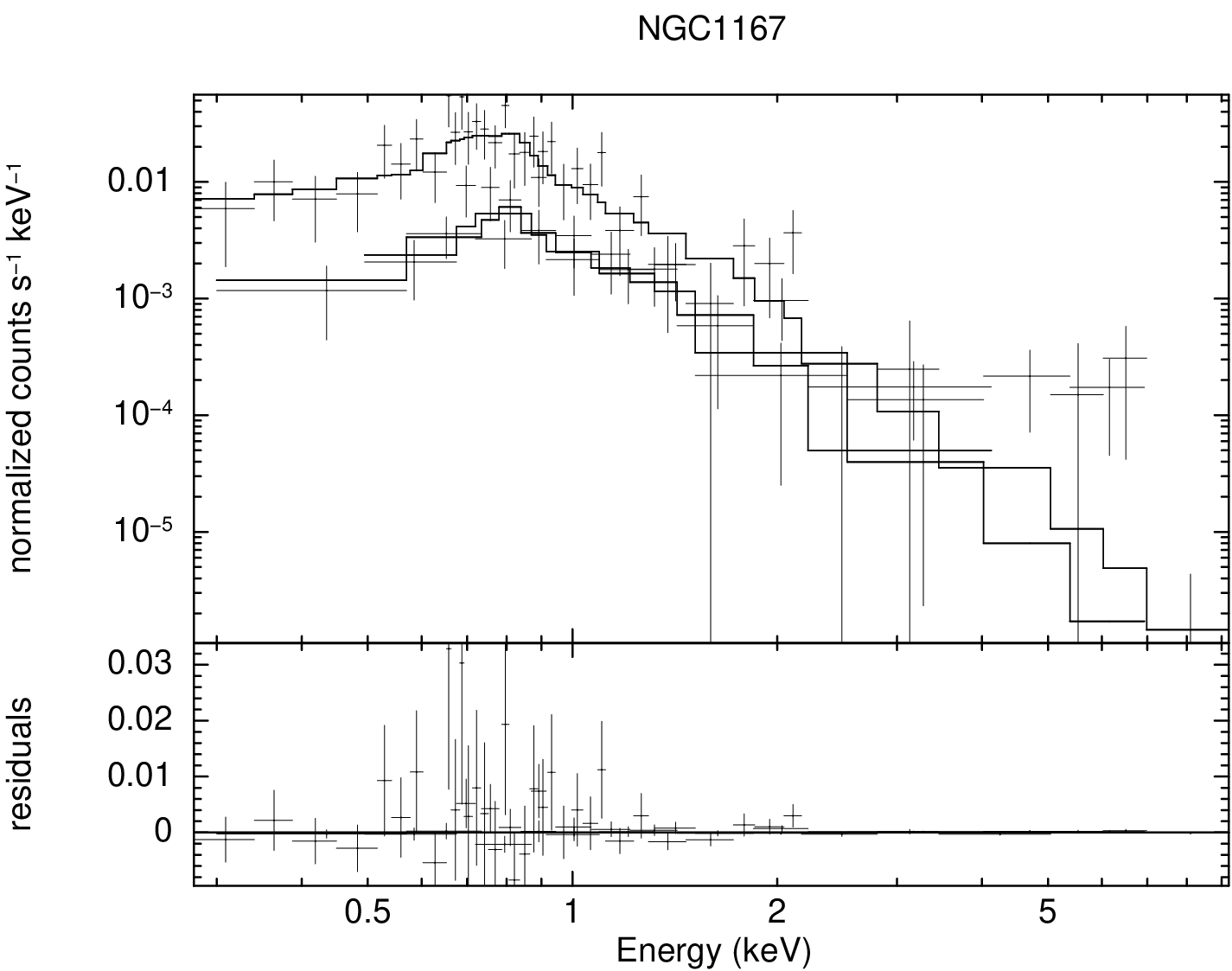}}
\rotatebox{-90}{\includegraphics[width=6cm, height=8cm]{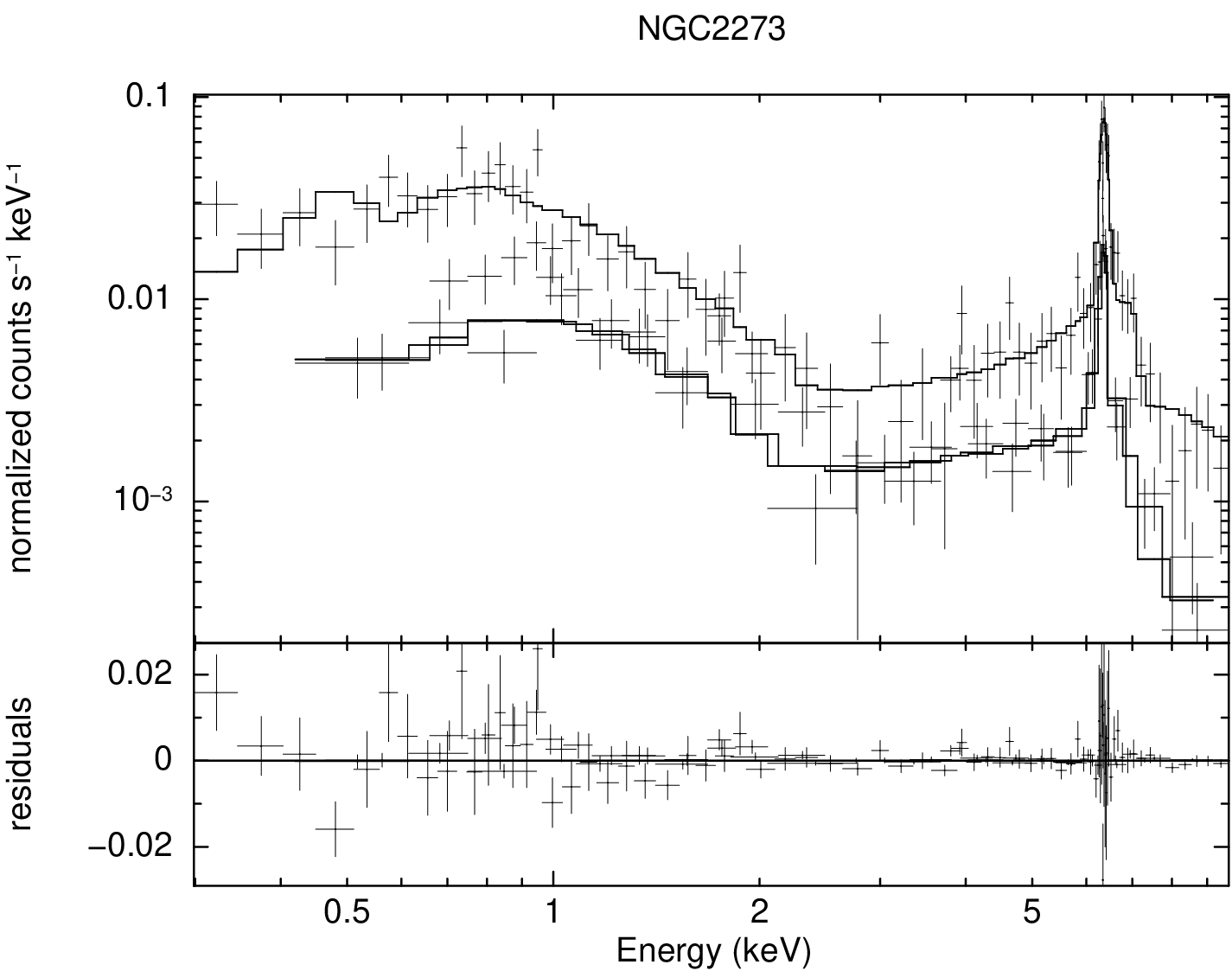}}
\rotatebox{-90}{\includegraphics[width=6cm, height=8cm]{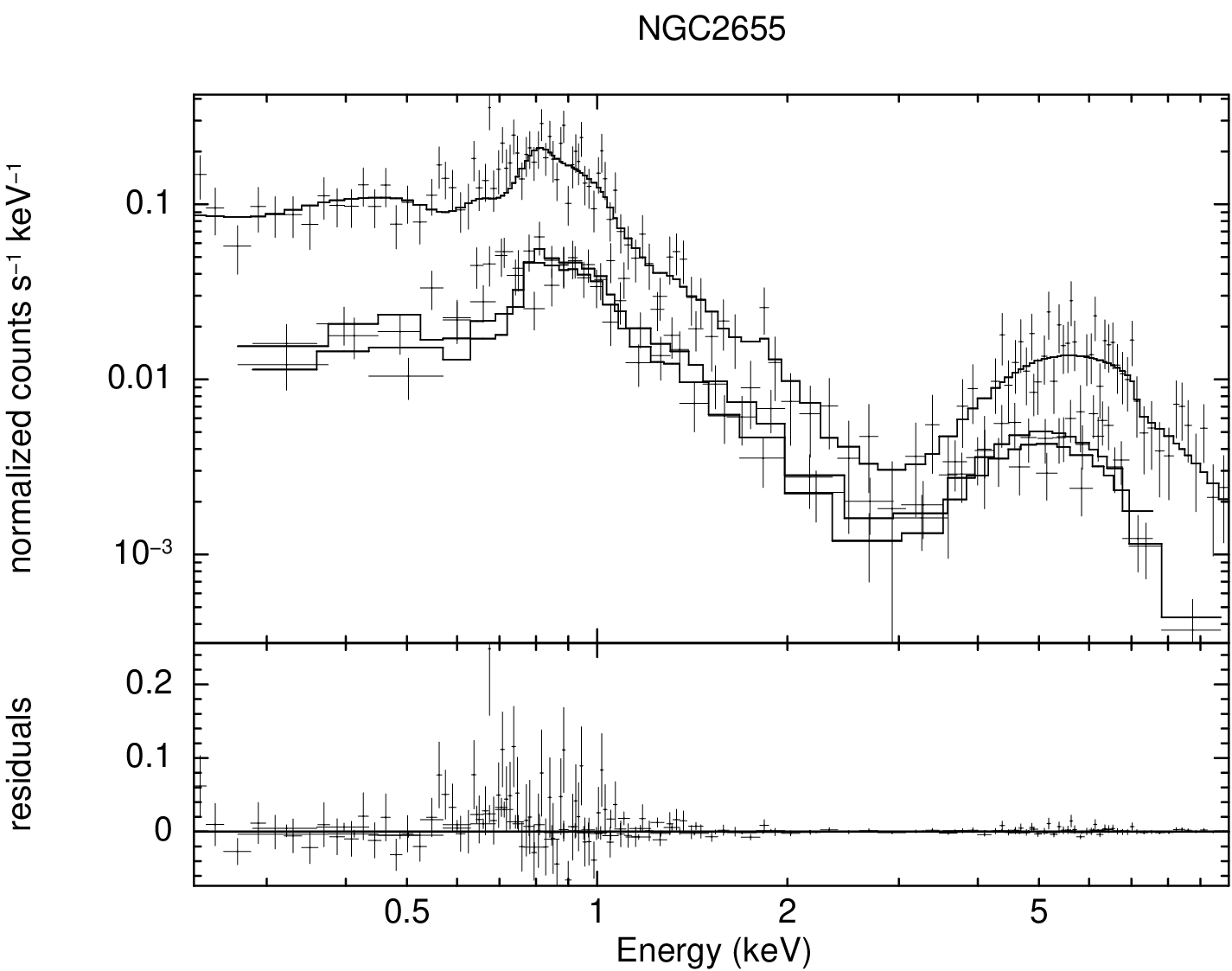}}
\rotatebox{-90}{\includegraphics[width=6cm, height=8cm]{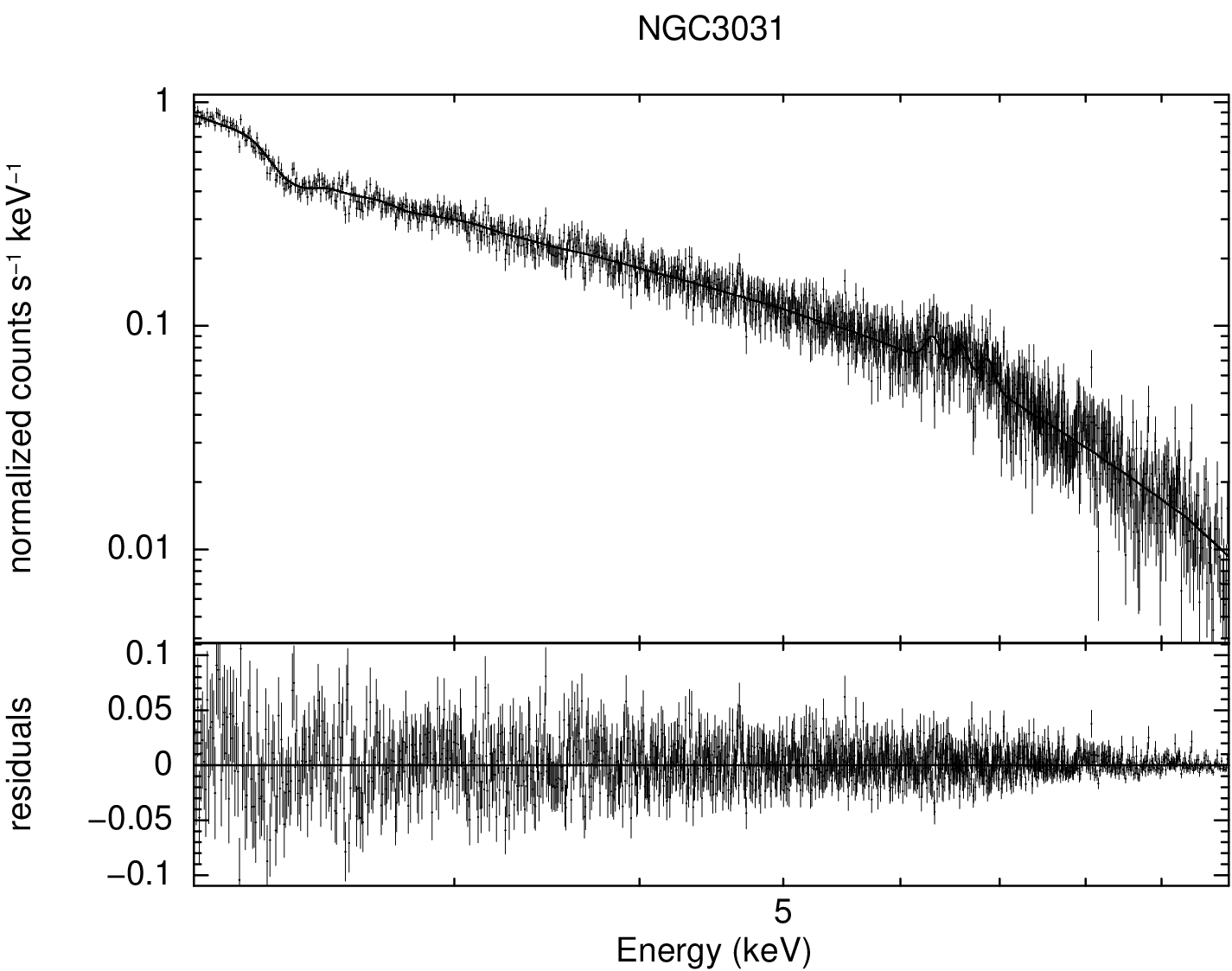}}
\rotatebox{-90}{\includegraphics[width=6cm, height=8cm]{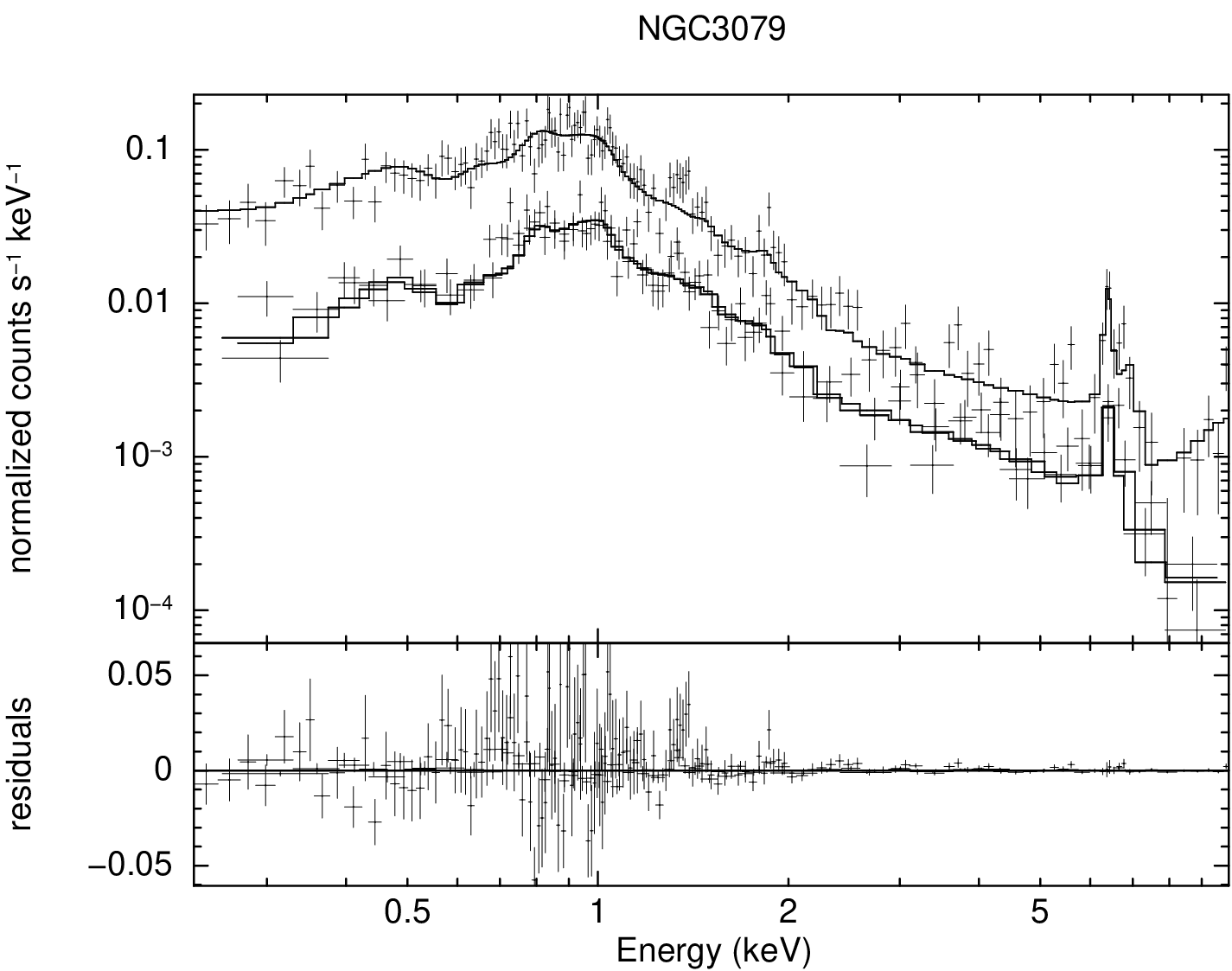}}
\rotatebox{-90}{\includegraphics[width=6cm, height=8cm]{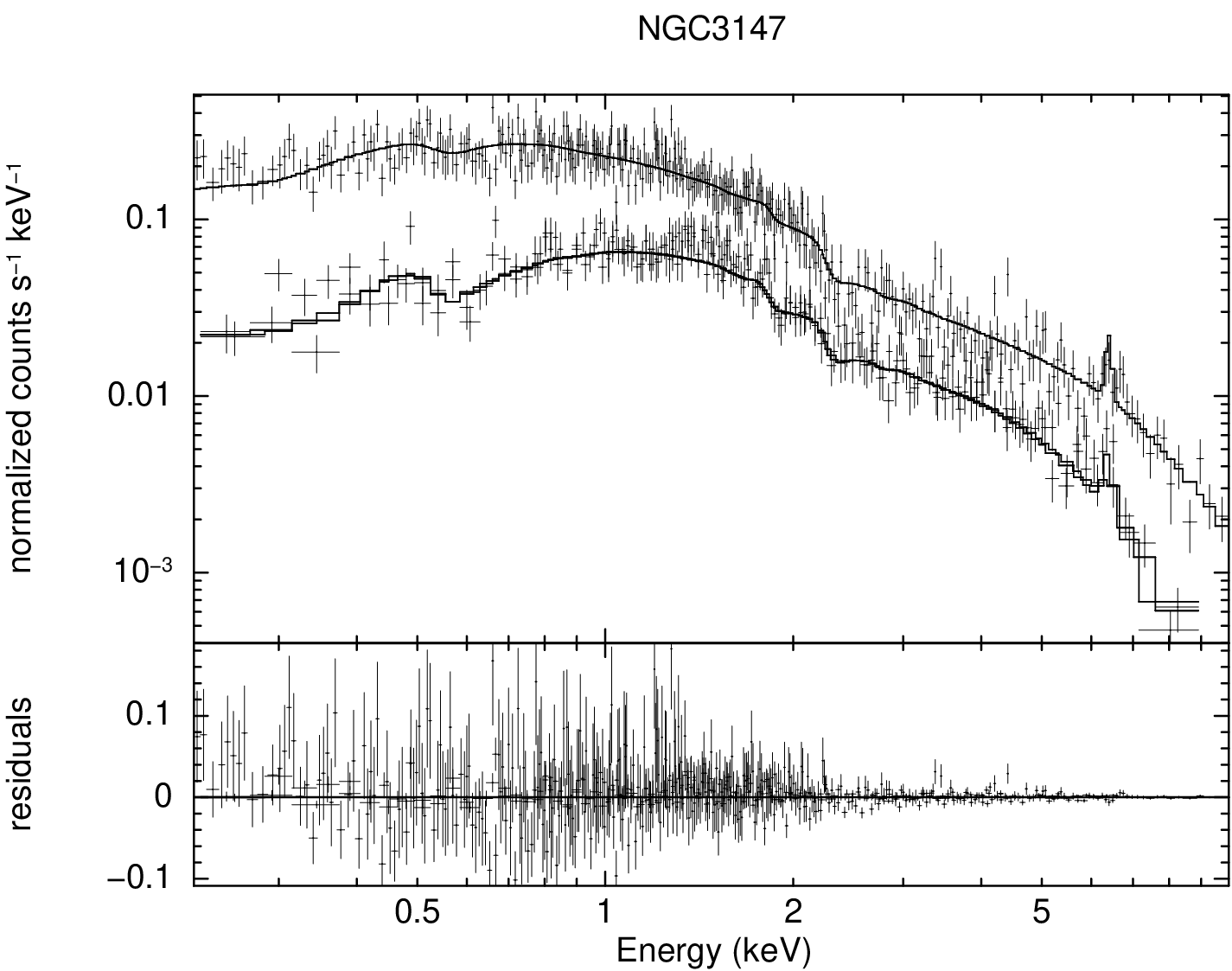}}
\label{xspectra}
\end{figure*}

\begin{figure*}
\centering
\rotatebox{-90}{\includegraphics[width=6cm, height=8cm]{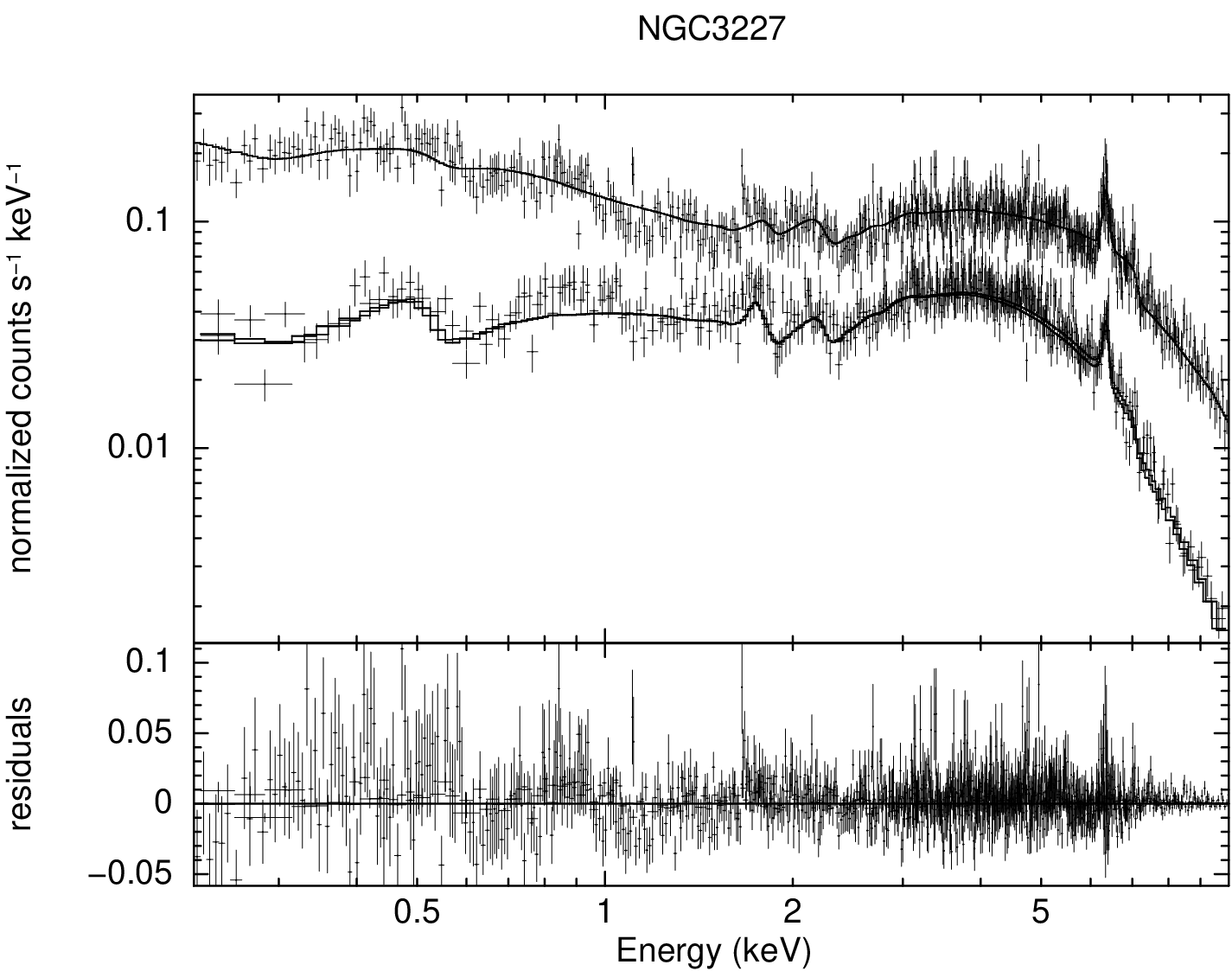}}
\rotatebox{-90}{\includegraphics[width=6cm, height=8cm]{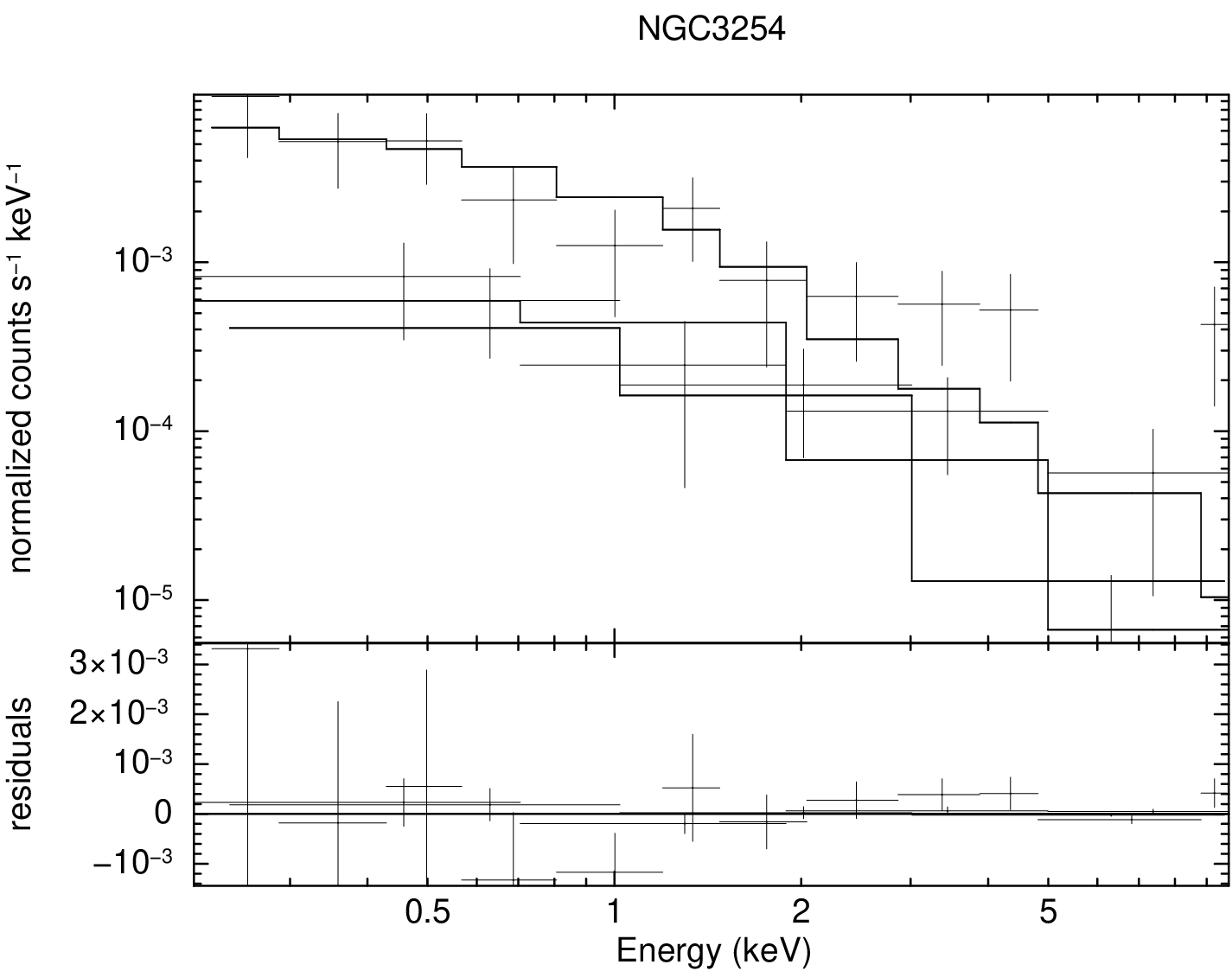}}
\rotatebox{-90}{\includegraphics[width=6cm, height=8cm]{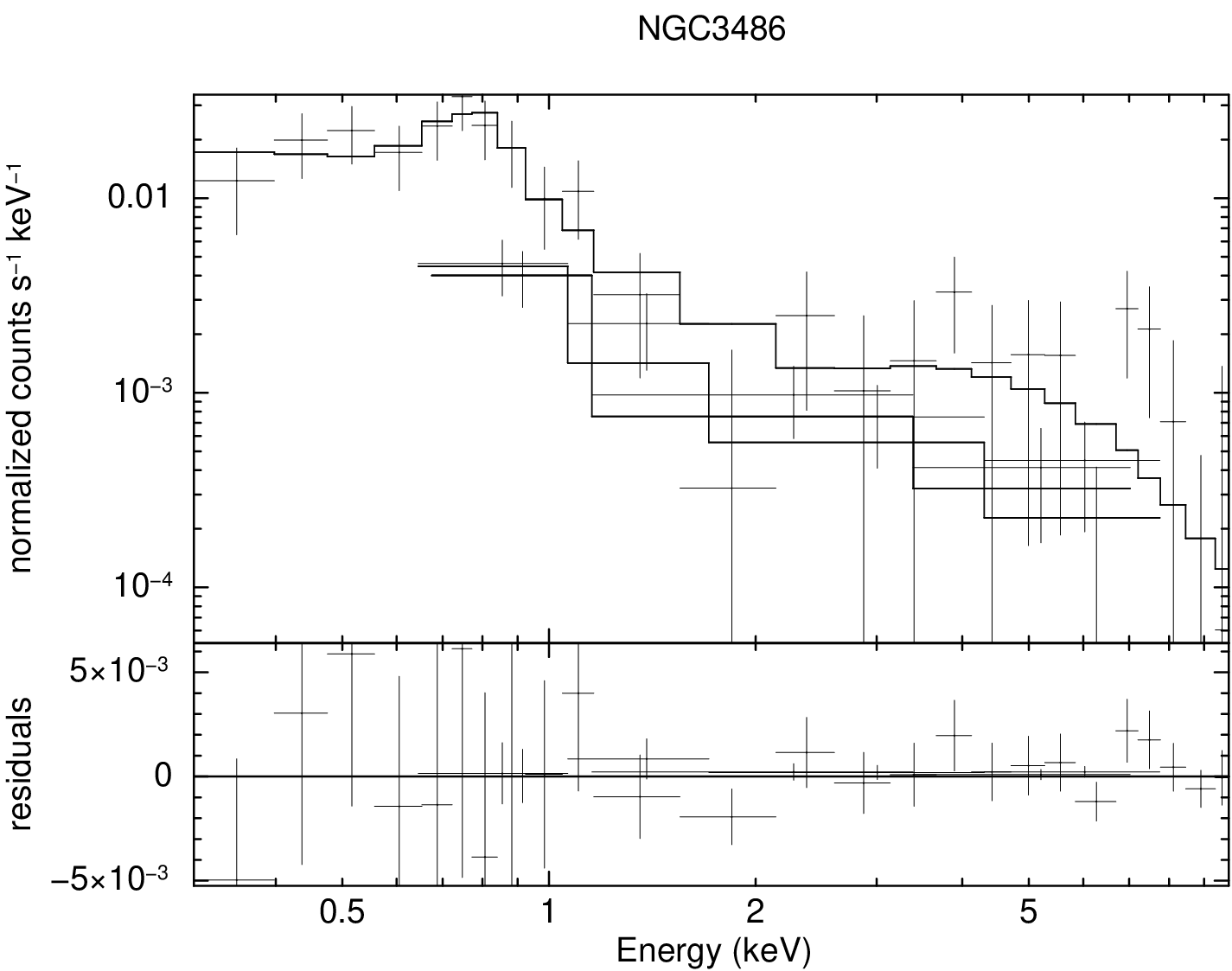}}
\rotatebox{-90}{\includegraphics[width=6cm, height=8cm]{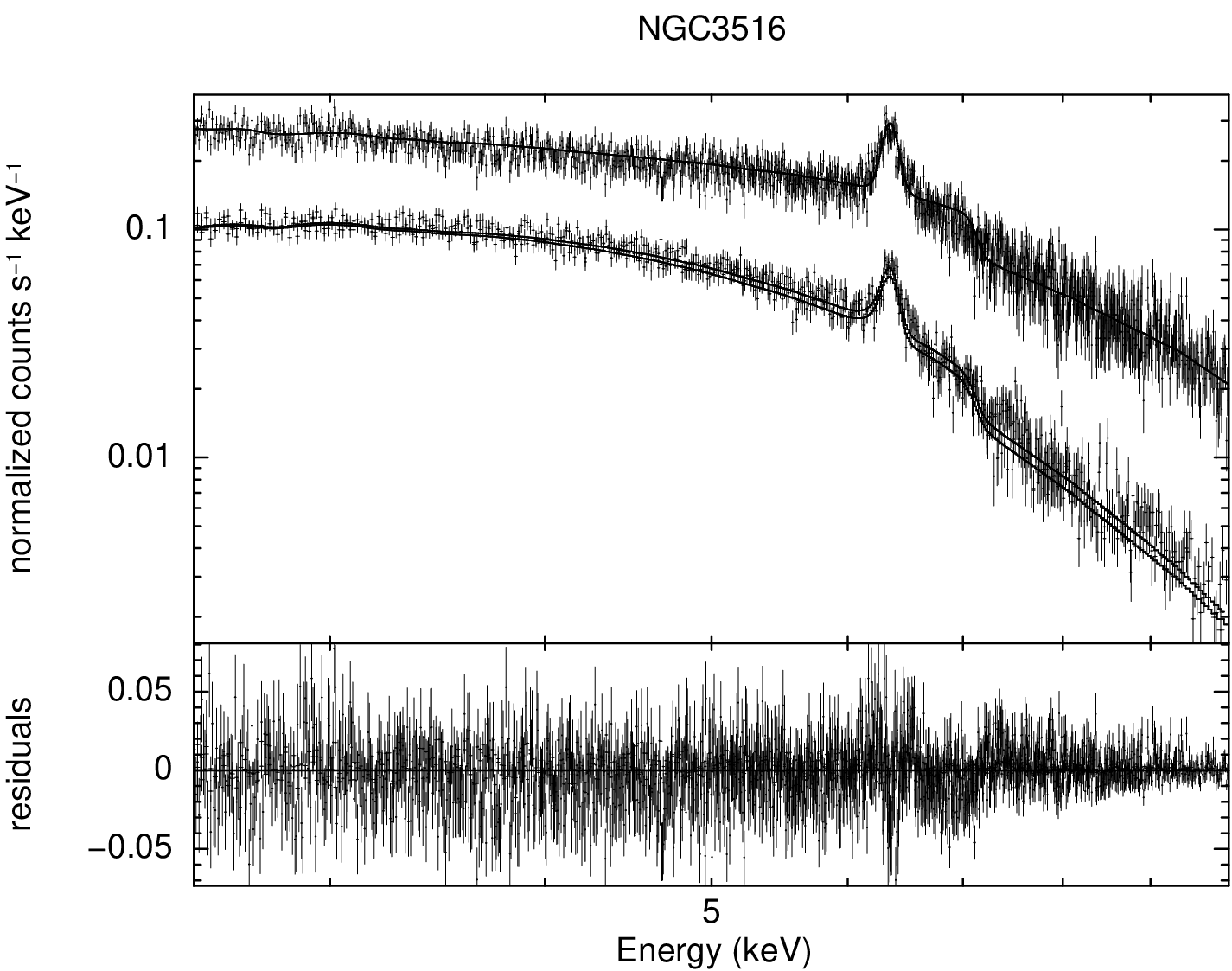}}
\rotatebox{-90}{\includegraphics[width=6cm, height=8cm]{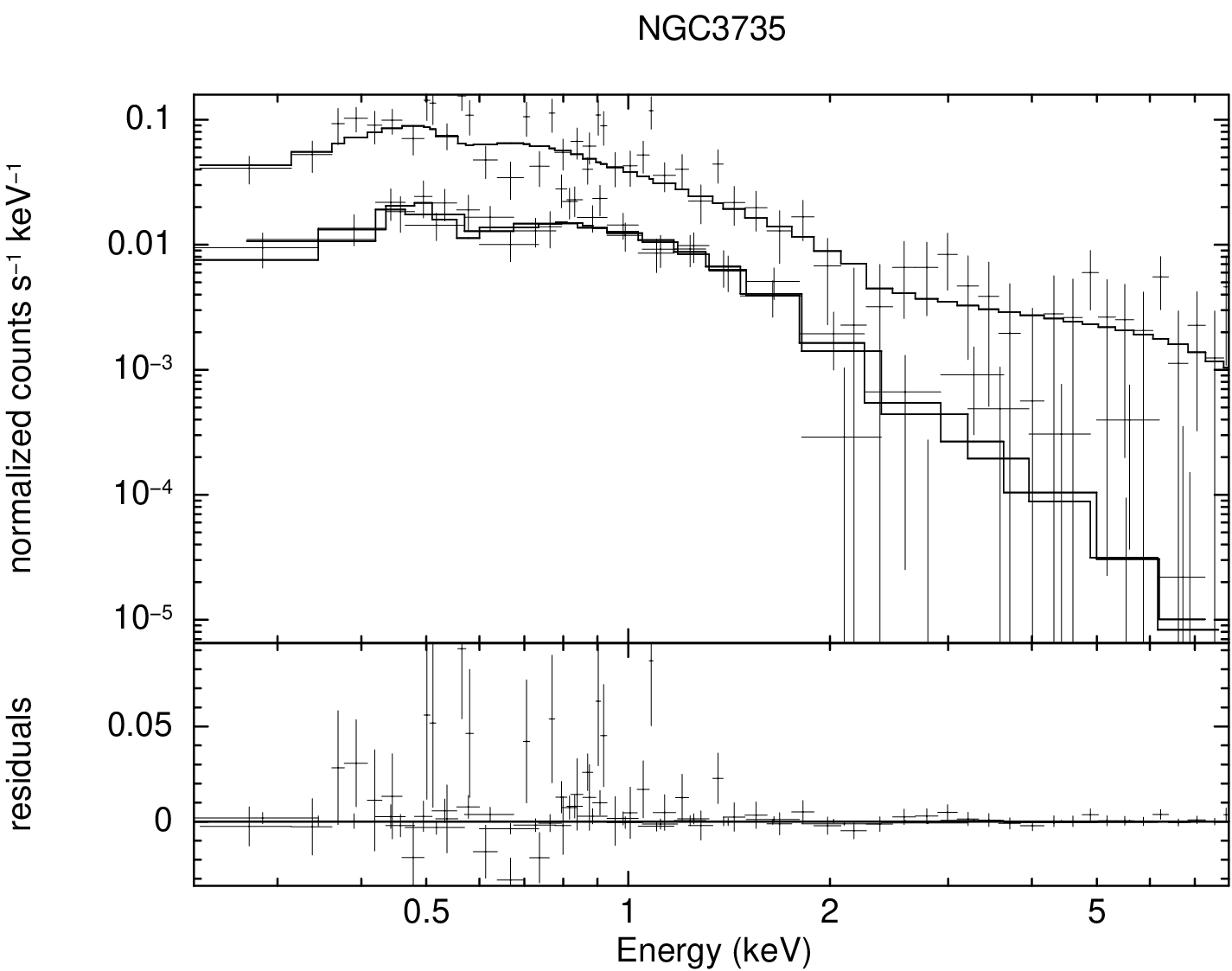}}
\rotatebox{-90}{\includegraphics[width=6cm, height=8cm]{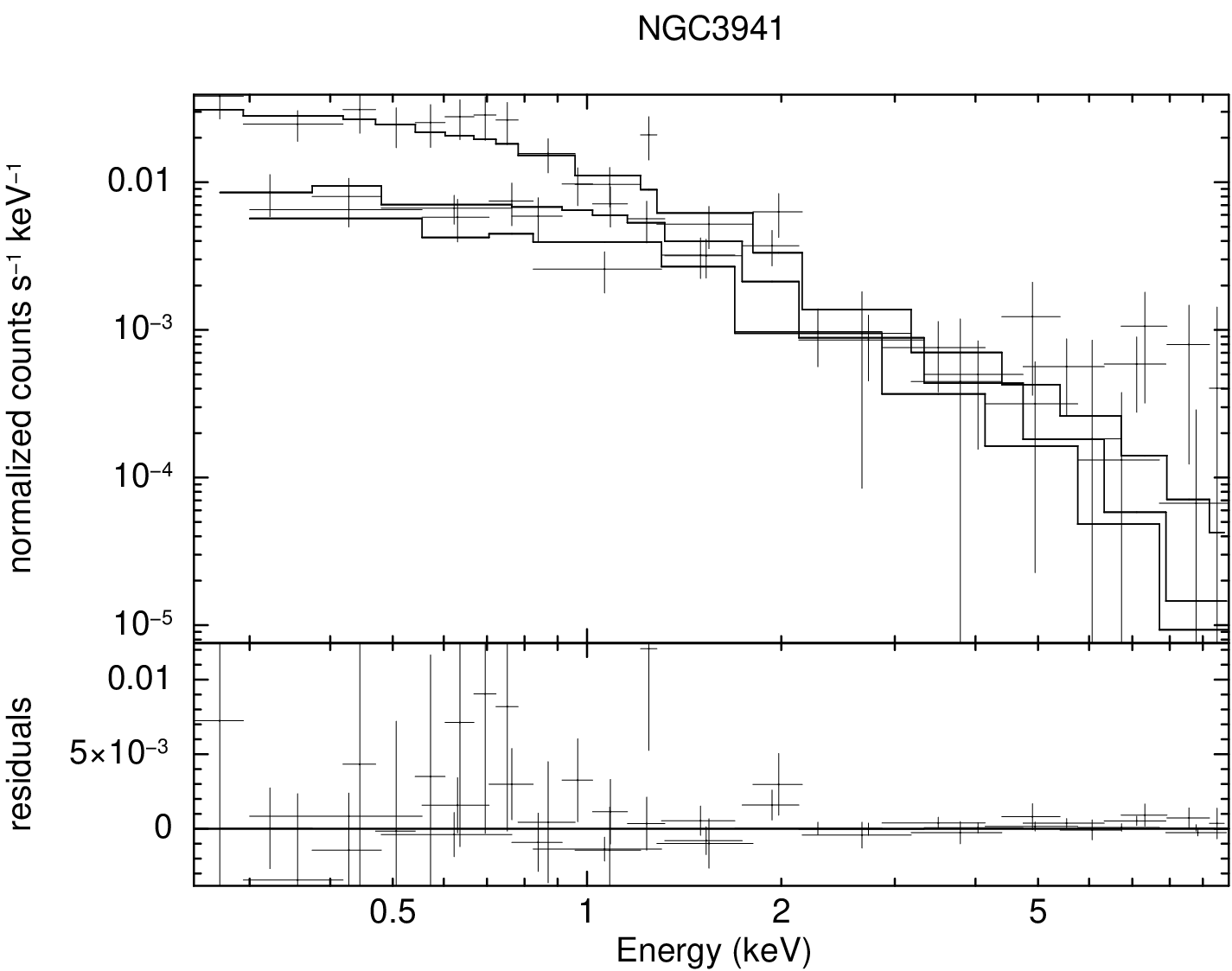}}
\rotatebox{-90}{\includegraphics[width=6cm, height=8cm]{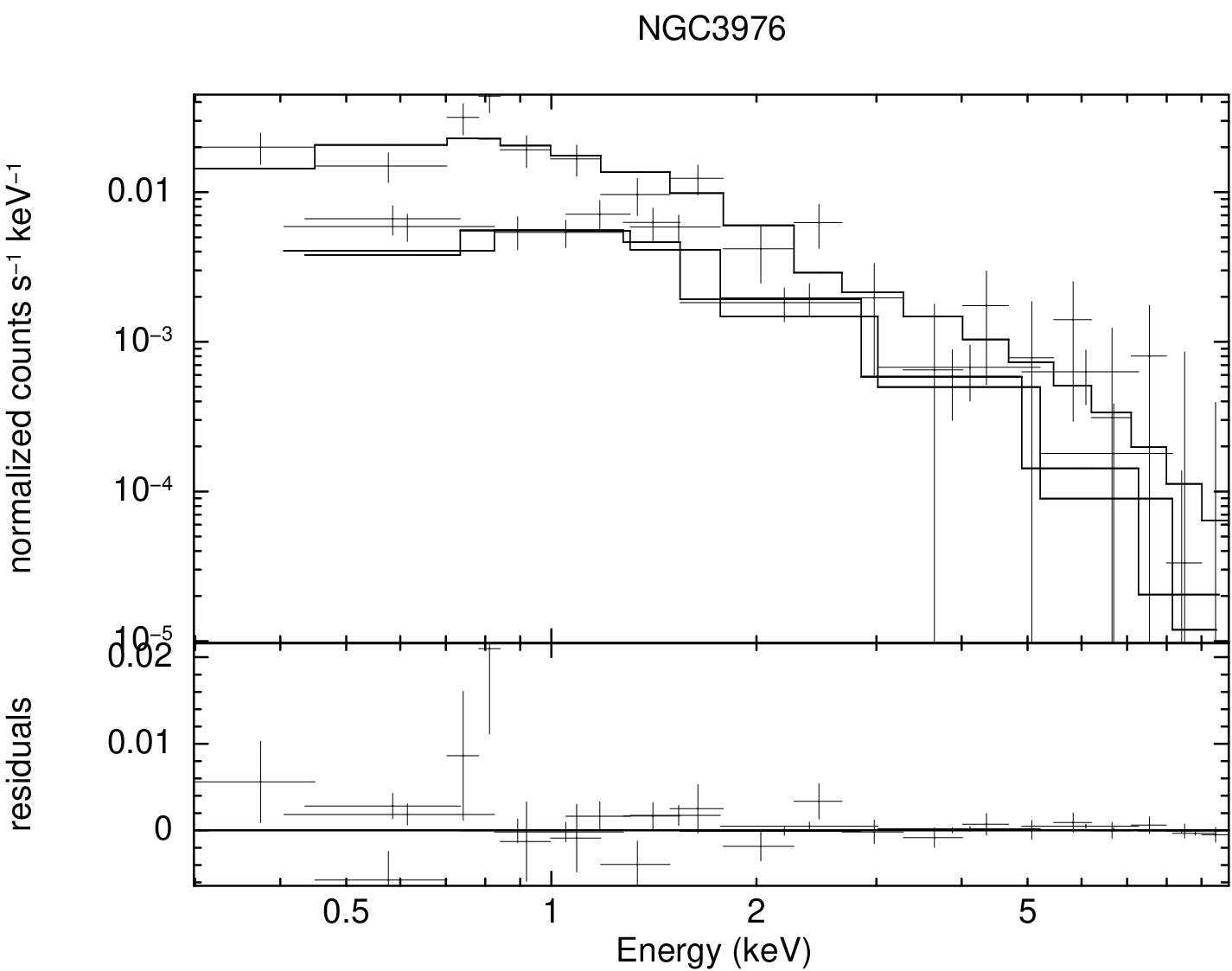}}
\rotatebox{-90}{\includegraphics[width=6cm, height=8cm]{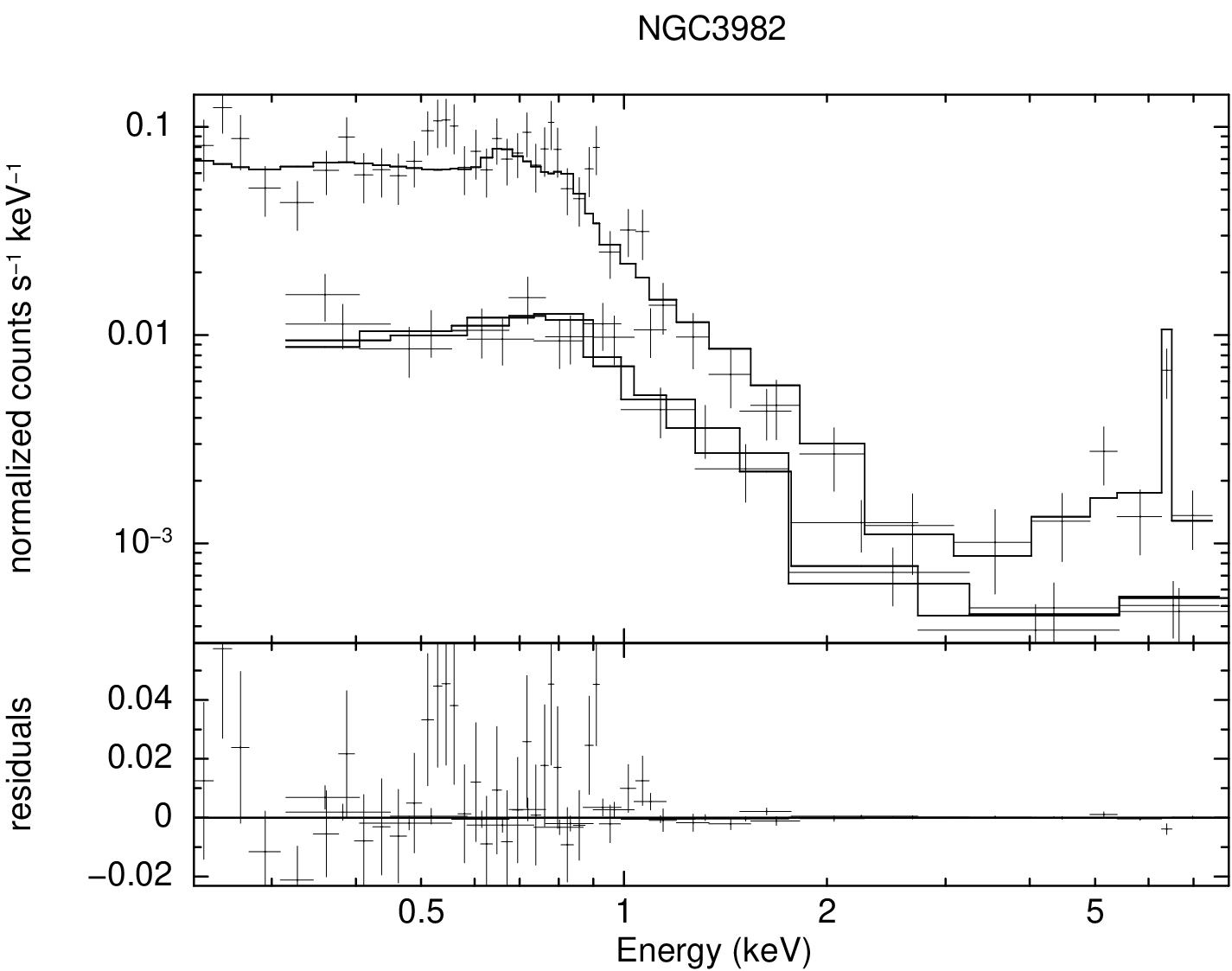}}
\label{xspectra}
\end{figure*}

\begin{figure*}
\centering
\rotatebox{-90}{\includegraphics[width=6cm, height=8cm]{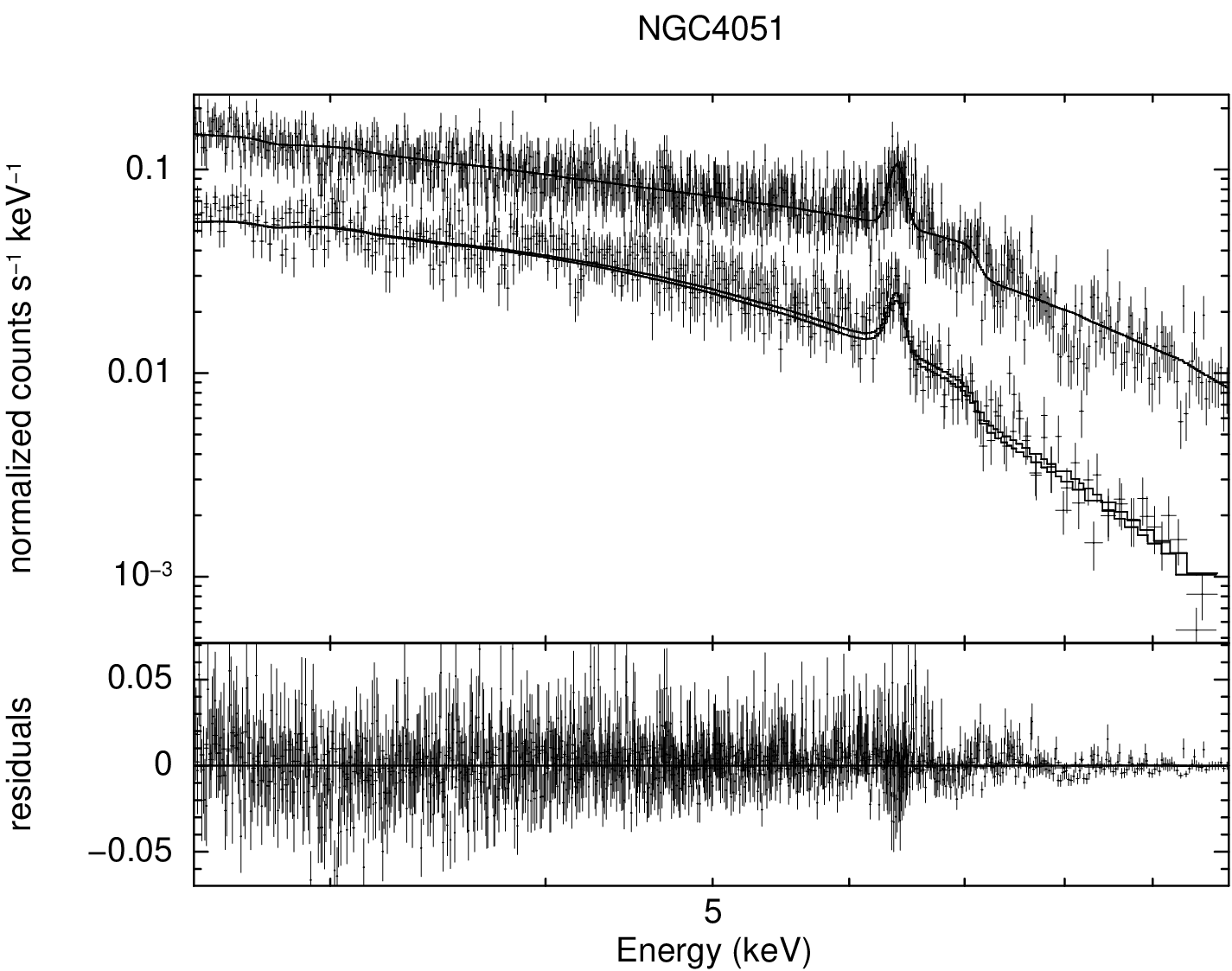}}
\rotatebox{-90}{\includegraphics[width=6cm, height=8cm]{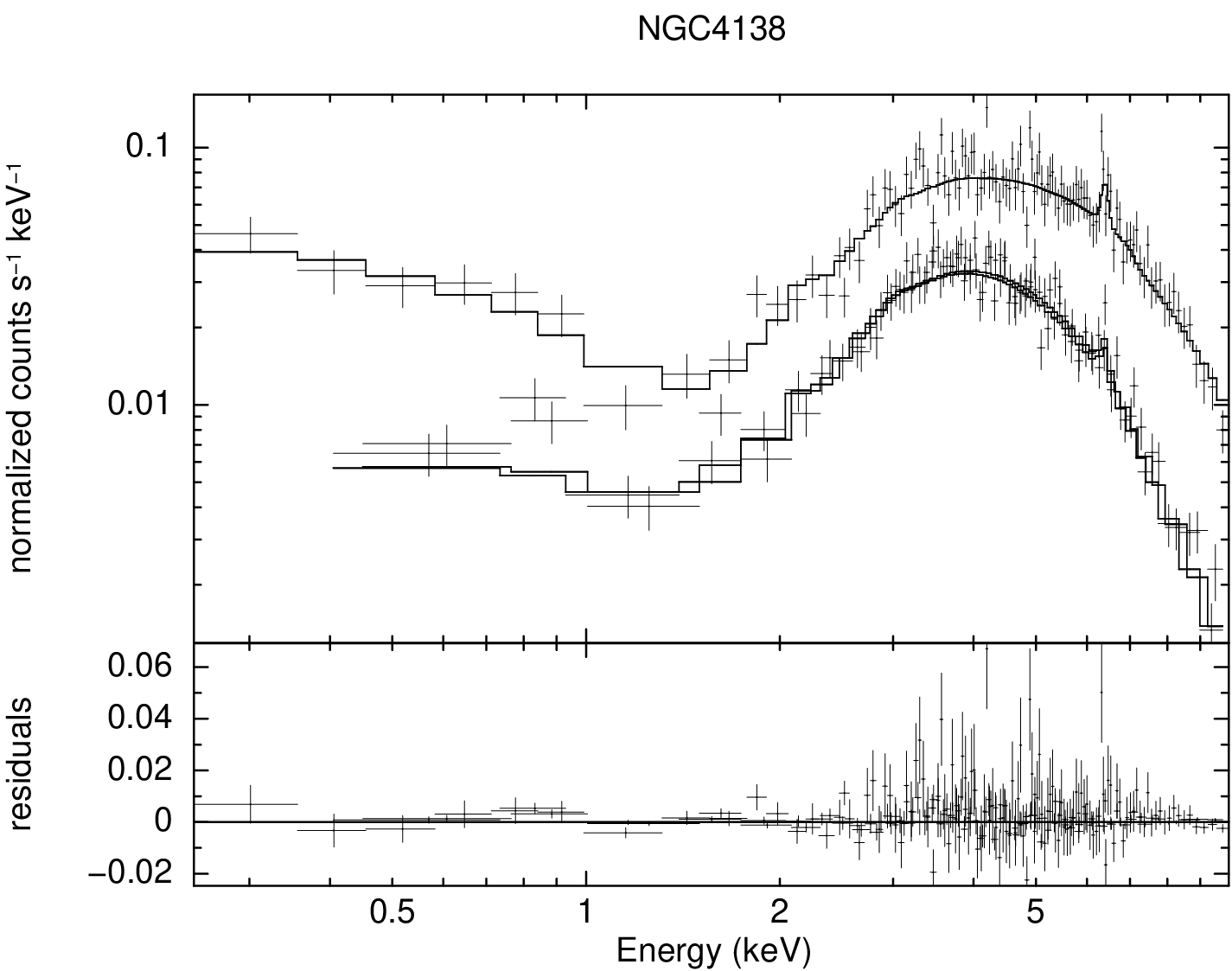}}
\rotatebox{-90}{\includegraphics[width=6cm, height=8cm]{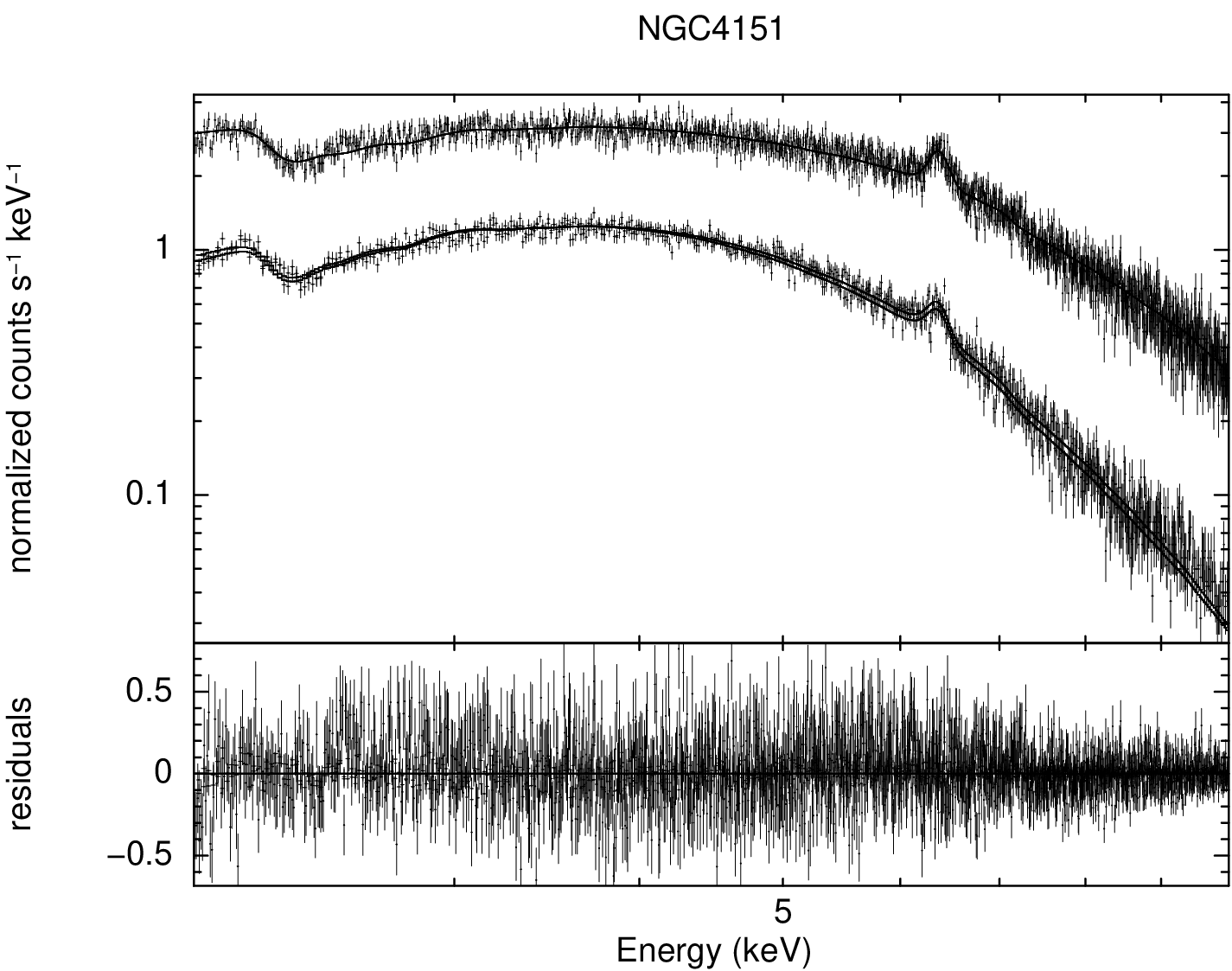}}
\rotatebox{-90}{\includegraphics[width=6cm, height=8cm]{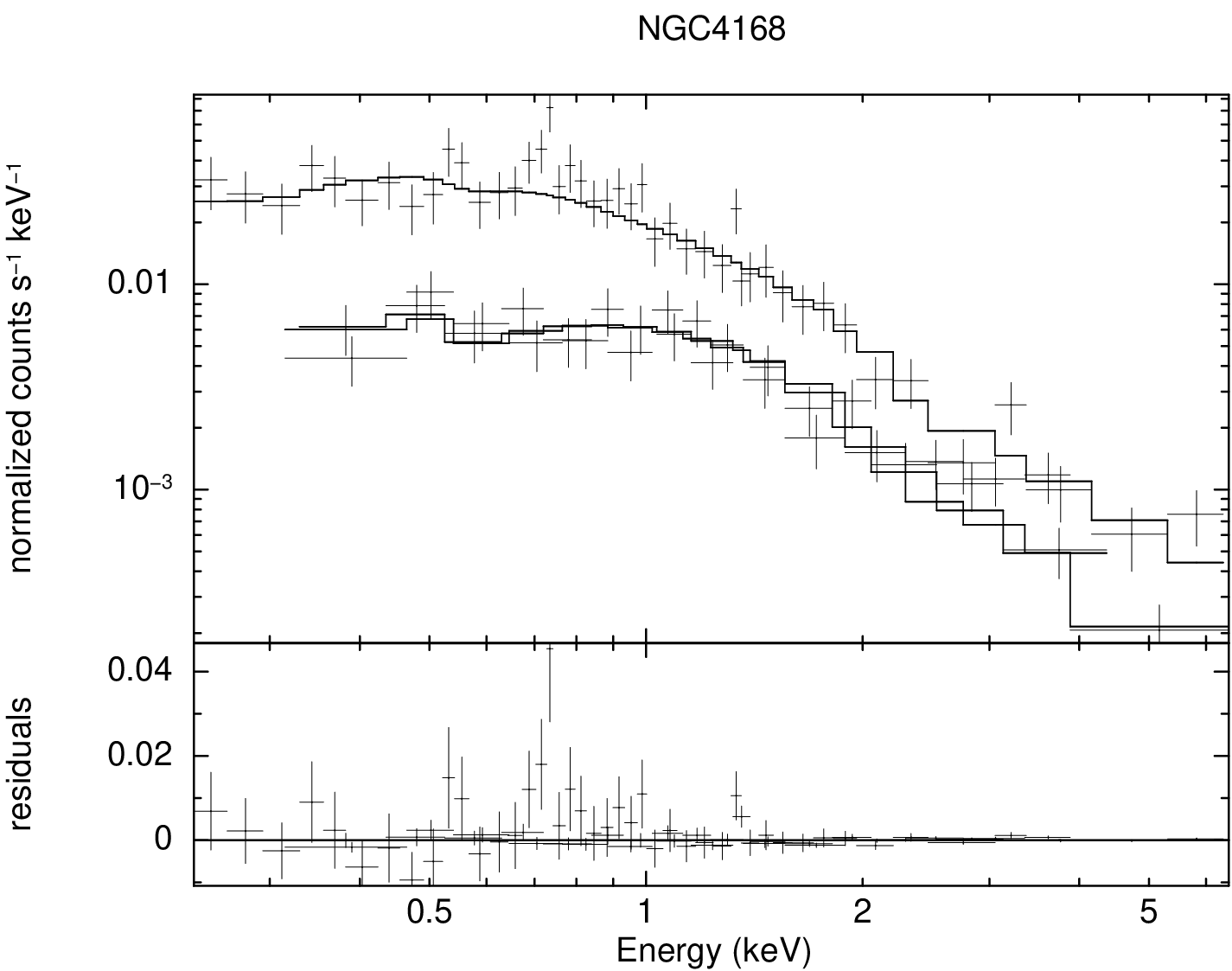}}
\rotatebox{-90}{\includegraphics[width=6cm, height=8cm]{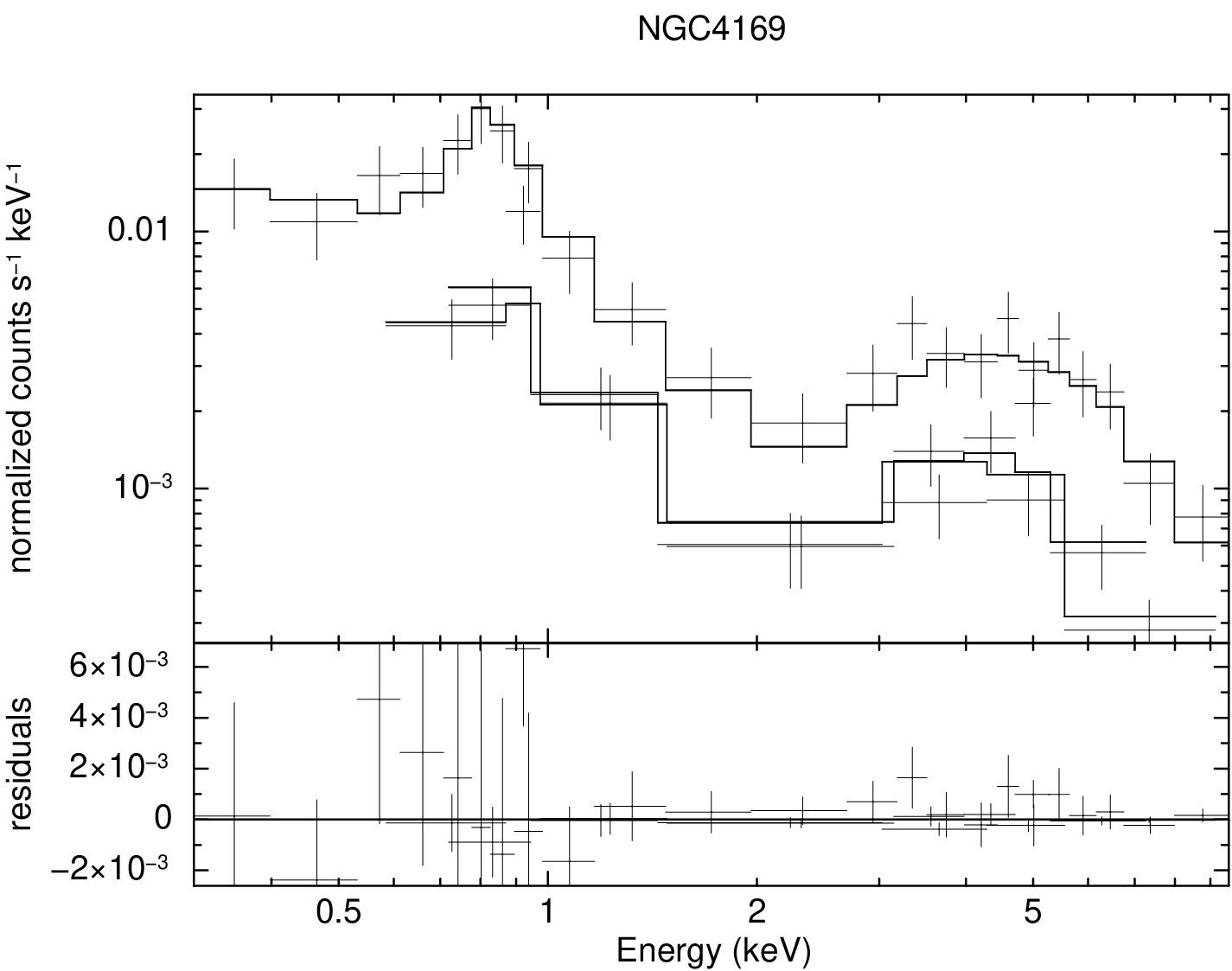}}
\rotatebox{-90}{\includegraphics[width=6cm, height=8cm]{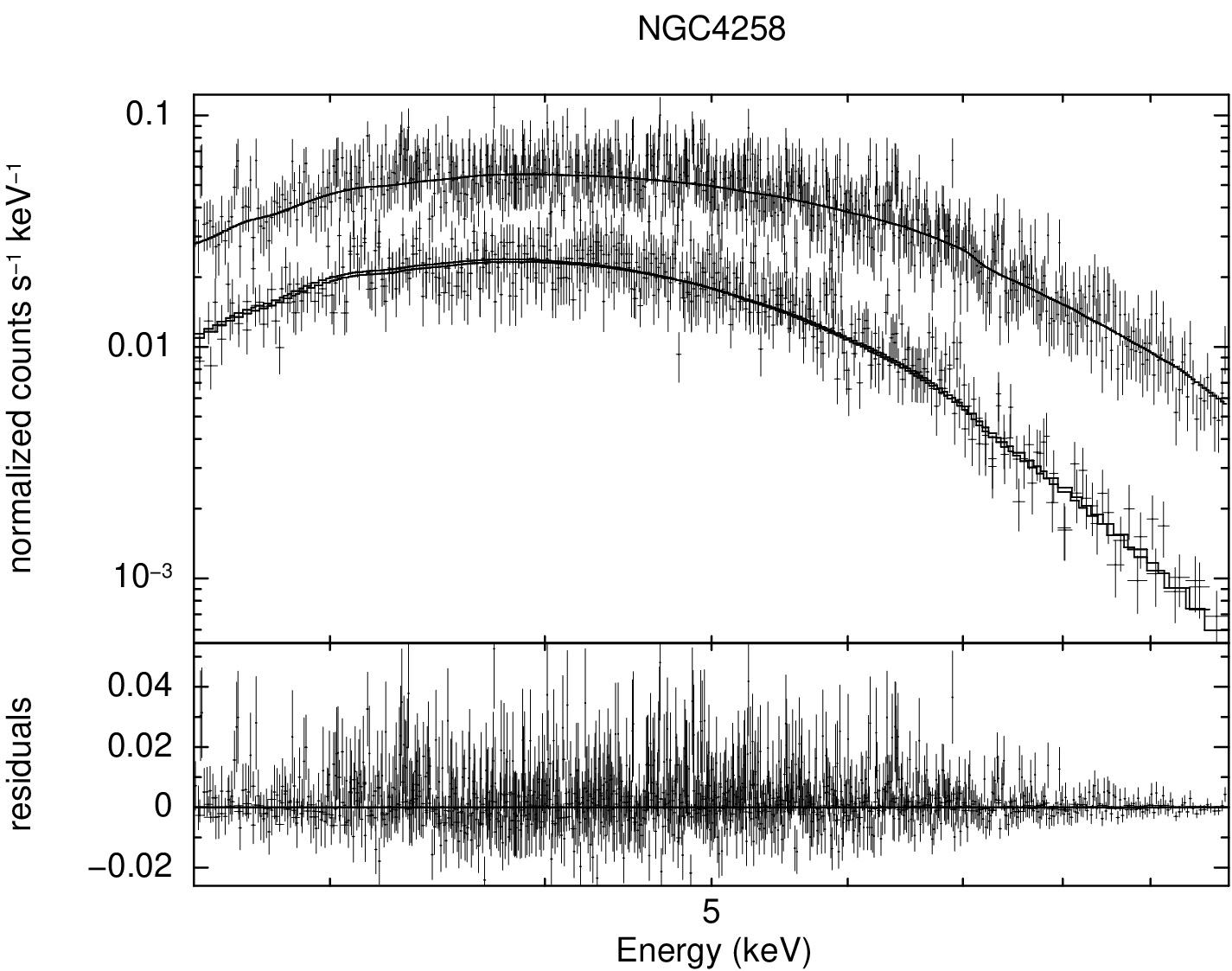}}
\rotatebox{-90}{\includegraphics[width=6cm, height=8cm]{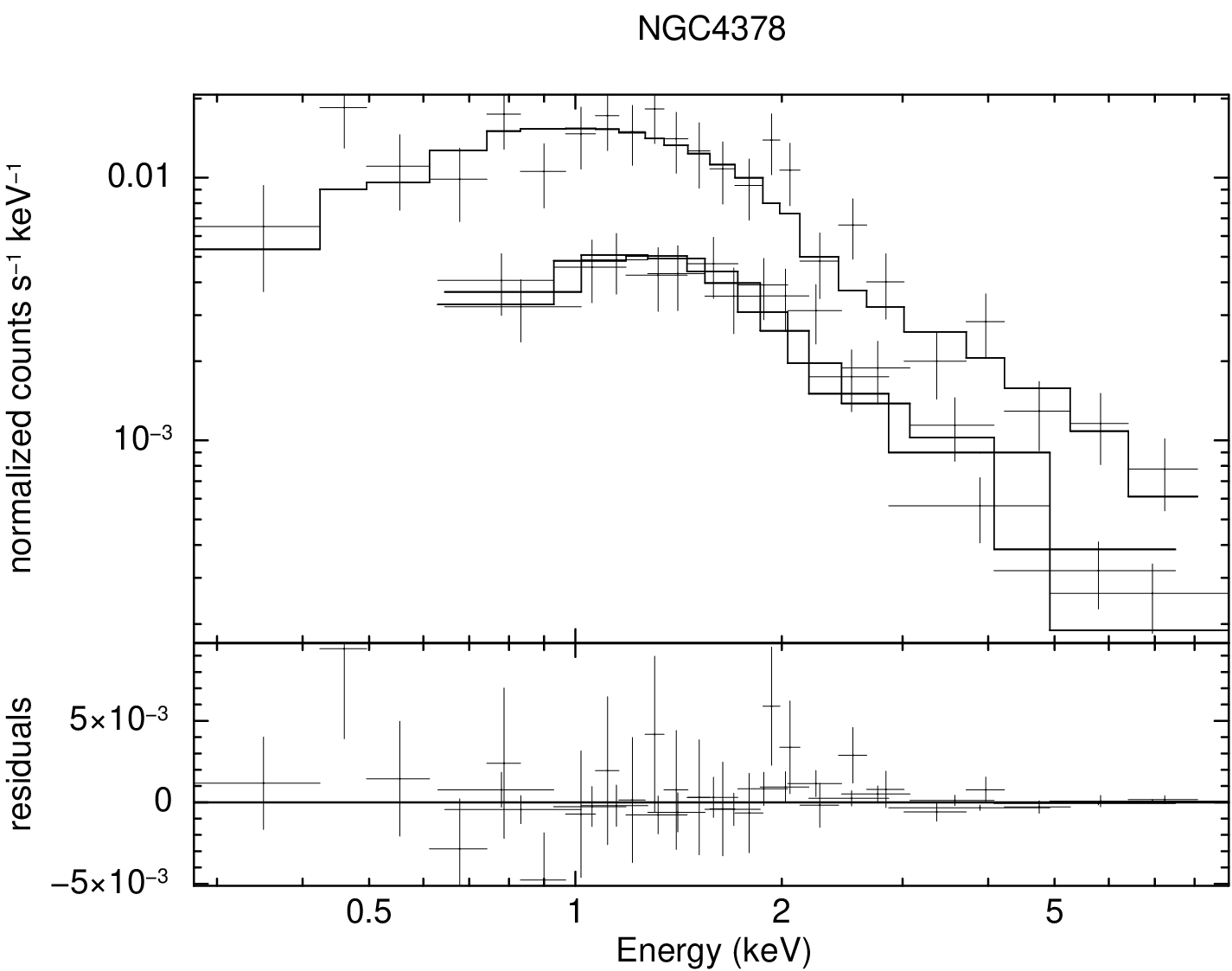}}
\rotatebox{-90}{\includegraphics[width=6cm, height=8cm]{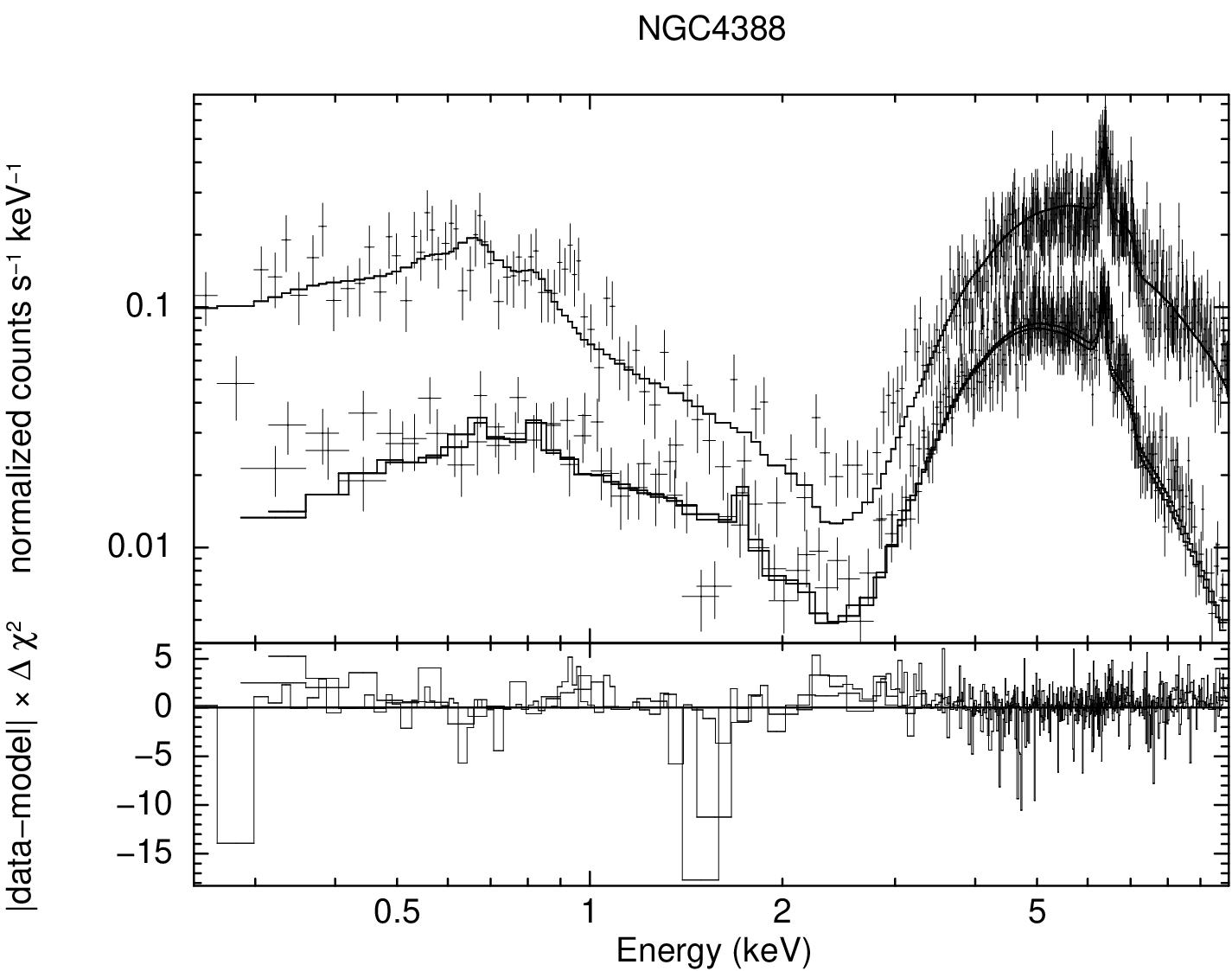}}
\label{xspectra}
\end{figure*}

\begin{figure*}
\centering
\rotatebox{-90}{\includegraphics[width=6cm, height=8cm]{4395.ps}}
\rotatebox{-90}{\includegraphics[width=6cm, height=8cm]{4472.ps}}
\rotatebox{-90}{\includegraphics[width=6cm, height=8cm]{4477.ps}}
\rotatebox{-90}{\includegraphics[width=6cm, height=8cm]{4501.ps}}
\rotatebox{-90}{\includegraphics[width=6cm, height=8cm]{4565.ps}}
\rotatebox{-90}{\includegraphics[width=6cm, height=8cm]{4639.ps}}
\rotatebox{-90}{\includegraphics[width=6cm, height=8cm]{4698.ps}}
\rotatebox{-90}{\includegraphics[width=6cm, height=8cm]{4725.ps}}

\label{xspectra}
\end{figure*}

\begin{figure*}
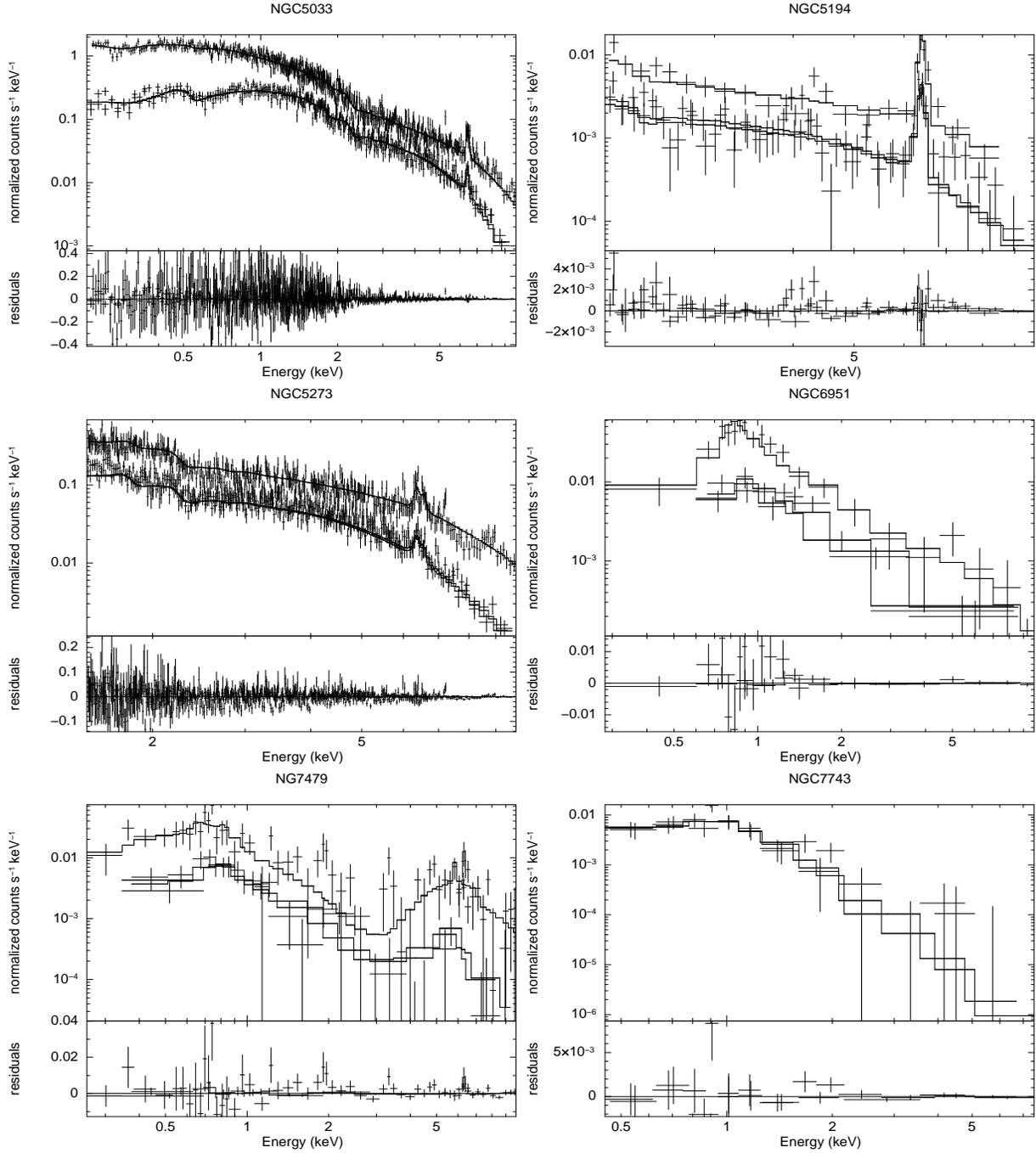

\centering
\rotatebox{-90}{\includegraphics[width=6cm, height=8cm]{5033.ps}}
\rotatebox{-90}{\includegraphics[width=6cm, height=8cm]{5194.ps}}
\rotatebox{-90}{\includegraphics[width=6cm, height=8cm]{5273.ps}}
\rotatebox{-90}{\includegraphics[width=6cm, height=8cm]{6951.ps}}
\rotatebox{-90}{\includegraphics[width=6cm, height=8cm]{7479.ps}}
\rotatebox{-90}{\includegraphics[width=6cm, height=8cm]{7743.ps}}

\caption{The \xmm X-ray spectra for all the sources in our 
sample. The upper panel shows the X-ray spectrum and the best fit model 
listed in Table \ref{fit} and the lower panel the residuals.}
\label{xspectra}
\end{figure*}

\end{document}